\documentclass{ws-ijmpd}
\usepackage{rotating}
\usepackage{graphicx}

\begin{document}

\markboth{Murat Hudaverdi et al.}
{Multi$-$band Analysis of Abell 2255}

%
\catchline{}{}{}{}{}
%

\title{MULTI$-$BAND ANALYSIS OF ABELL 2255
}

\author{Murat Hudaverdi
, E. Nihal Ercan, Arzu Mert-Ankay}

\address{Bo\u{g}azi\c{c}i University, Physics Department\\80815 Bebek, Istanbul, Turkey\\ 
murat.hudaverdi@boun.edu.tr}

\author{Fatma G\"{o}k, Ebru Aktekin}
\address{Akdeniz University, Physics Department\\ Dumlupinar Boulevard, 07058,
  Antalya, Turkey}

\author{G\"{u}lnur \.{I}kis G\"{u}n, Burak U\u{g}ra\c{s}}
\address{18 Mart \c{C}anakkale University, Physics Department\\ Terzioglu Campus, 17100, Canakkale, Turkey\\}

\author{Tolga G\"{u}ver}
\address{Istanbul University\\ Astronomy \& Space Sciences Department\\ 34119, Istanbul, Turkey}

\author{Hideyo Kunieda}
\address{Nagoya University, Department of Physics\\ Furo-cho, Chikusa, 464-8602. Nagoya, Japan}


\maketitle

\begin{history}
\received{-- -- ----}
\revised{-- -- ----}
\comby{Managing Editor}
\end{history}

\begin{abstract}
We analyzed the data for the nearby cluster of galaxies Abell 2255 from
archival {\em XMM-Newton}
observations in order to search for the properties of X-ray point like structures in the outskirts.
11 point-like X-ray emission is detected.
Detected X-ray sources are then observed with the 1.5 meter 
{\em RTT-150} at Turkish National Observatory for possible optical counterparts.
The cluster field is covered through 5 {\em ANDOR} photometer pointings.
3 sources have no optical follow-ups.
2 QSOs and 1 star are observed from the field.
For 4 sources we have obtained the corresponding redhifts.
The cumulative log($N$)-log($S$) is studied and the cluster source number is
calculated to be 4 times higher than the field at
$F_x$$\sim$10$^{-13}$ ergs cm$^{-2}$ s$^{-1}$. 
This phenomenon is interpreted as increased galaxy activity as they first encounter
high density ICM environment at the cluster outskirt.
We suggest that X-ray emission is triggered by either increased accretion onto
LMXBs, fueling of AGNs and/or awakening of BHs.

\end{abstract}


\section{Introduction}	
Clusters of galaxies are the largest aggregates of galaxies.
They are formed from the gravitational collapse of the field galaxies and the
subgroups which is a recurrent event and still ongoing at the present epoch.
Samples of galaxies extracted from a certain cluster are subject to the same
selection effects because they all accomodate at the same redshift. 
Cluster centers come to dynamical equilibrium prior to outer regions.
Consequently the substructral features vanish in the core.
Whereas the cluster outskirts are dynamically complex regions populated by
galaxies moving towards the cluster potential.
Understanding the undergoing physics in these regions between the virialized
cluster cores and the widespread field is specially important.

Although early optical surveys suggest that active galaxies in clusters are
relatively lower than the field (Refs. \refcite{ost-60}; \refcite{gis-78}; \refcite{dre-85})
, the conventional portrait of clusters shows overdensities of galaxies as expected. 
The recent X-ray results (Refs. \refcite{cap-01}; \refcite{mol-02}; \refcite{joh-03}) 
on {\em Chandra} and {\em XMM-Newton} data reveal
higher fractions of point sources toward clusters of galaxies compared to
blank fields. Refs. \refcite{cappe-05} and \refcite{rud-05}
conducted systematic analysis of X-ray source populations and found
significant excess from cluster fields.
As data for more clusters become available, it is concluded that the brighter
galaxies are likely to locate in the outskirts of clusters ({\em R} $\sim$ 1 Mpc), which is naturally
attributed to the infall-related fueling of active nuclei
(See Refs. \refcite{mar-06}; \refcite{hud-06}).  

In order to understand environmental effects of high-density Intra-cluster medium (ICM) on galaxies, 
we selected a dynamically active merging system, 
dominated by point-like structures not buried into the ICM diffuse emission.
In this paper, we report X-ray and optical observational results of galaxies
from the outskirts of Abell 2255.
A2255 is a nearby ($z$=0.086), bright cluster showing several signs of a merger event. Based on optical
data, Ref. \refcite{yua-03} reported subgroups of galaxies in the
outskirts, rotating around the center of A2255. Two bright central galaxies
which are the survived central-dominant (cD) ellipticals of merged subclusters
and high velocity dispersions (Ref. \refcite{bur-95}) are the other evidences of large scale dynamics. 
A2255 is also bright in radio wavelenghts. 
The cluster contains a central radio halo and relics (Refs. \refcite{gio-99}; \refcite{gov-05}). 
It is well known fact that radio emission comes from relativistic electrons 
which can be accelerated by large scale cluster mergers.
A2255 is also well studied in X-rays. {\em ROSAT} data results showed that X-ray peak
is offset from the brigtest galaxy (Refs. \refcite{dav-95}; \refcite{bur-95}). More recently,
Ref. \refcite{sak-06} observed A2255 by {\em XMM-Newton} and reported
temperature asymmetries, indicating subunit mergings. 
Ref. \refcite{dav-03}, analyzed {\em Chandra} data to investigate
the X-ray properties of cluster galaxies in the central ({\em R} $\le$ 0.5
Mpc) region. The study confirms the results of Ref. \refcite{mar-02} that
X-ray selected AGN fraction is higher than the optically selected AGNs and
these AGNs likely to reside in red early-type galaxies.

The purpose of this paper is to study source properties in the cluster
outskirts with X-ray and optical properties. 
We present the data in $\S$ \ref{data}. The following section $\S$
\ref{analysis} describes the analysis steps.
The results are discussed in $\S$ \ref{discussion}.
Assuming $H_o$=75 km s$^{-1}$ Mpc$^{-1}$ and cosmological deceleration parameter of $q_0$=0.5,
the luminosity distance to A2255 is found to be 348 Mpc, and an angular size of 1 arcsec corresponds to 1.63 kpc.
The quoted uncertainties for the best fit parameters of spectral fittings in
$\S$3.1 are given for 90\% confidence range.

\section{\label{data}Observations and Data Reduction}
\subsection{X-Ray Data}
We analysed archival X-ray data ({\em PI: Turner, M.}) obtained by two {\em XMM-Newton} observations. 
Both of the observations are covering entire cluster emission with $\sim$3 arcmin offset from the X-ray peak. 
The three EPIC instruments, the two MOSs (Ref. \refcite{tur-01}) and the PN (Ref. \refcite{str-01}), were used. 
A2255 was observed in revolutions 525 and 548. 
The cameras were operated in the Prime Full Window mode for MOS and Prime Full Window Extended mode for PN. 
The thin filter was used for all EPIC cameras.  
We processed the observation data files and created calibrated event files using the SAS version 7.0.0. 
The event lists were generated from the observation data files (ODF) by the
tasks {\scriptsize EMCHAIN} and {\scriptsize EPCHAIN}. 

\begin{table}[h]
\centering
\tbl{Log of {\em XMM-Newton} observations.}
{\begin{tabular}{@{}cccccc@{}} 
\toprule
 OBS-ID  &  Date      & \multicolumn{3}{c}{Exp$_{(live)}$/Exp$_{(corr)}$ (ks)} & Filter \\
                         &            &     MOS1  &    MOS2    &    PN                 &  EPIC  \\ 
\colrule
 0112260501  & 2002.10.22 &     8.73 / 6.76    &  8.82 / 6.80  & 5.98 / 3.11   & thin   \\
 0112260801  & 2002.12.07 &    16.69 / 9.60    &  16.65 / 9.82 & 17.39 / 7.40
& thin   \\ 
\botrule
\end{tabular} \label{table1}}
\end{table}

\begin{figure}[htb]
 \begin{center} 
\includegraphics[width=8cm]{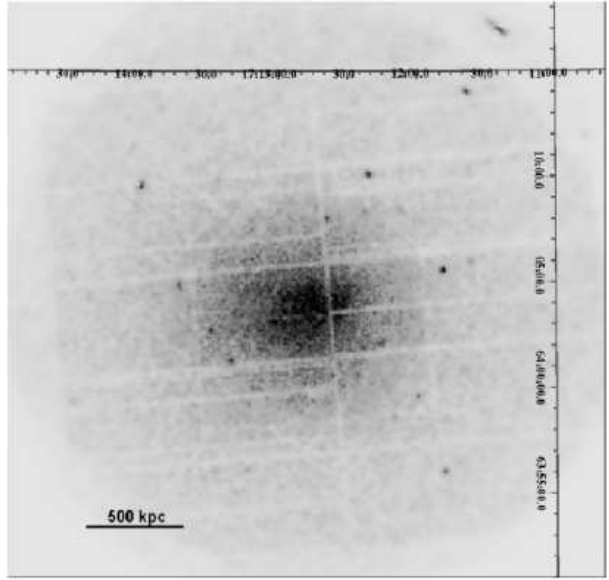}
   \end{center}
  \caption{Adaptively smoothed XMM-Newton merged EPIC image of A2255 in the
    0.3-10 keV band. The image has a size of 30$^{\prime}\times$30$^{\prime}$ and
    a pixel size of 5$^{\prime\prime}$. It is adjusted with north up
    and east left. The bar shows 500 kpc scale size. 
    The circles indicate the x-ray source locations. 
}\label{fig:A2255_with_sources}
\end{figure}

In the standart analysis, the events were selected with {\scriptsize PATTERN 0-12} for
MOS and single \& double pixel events ({\scriptsize PATTERN 0-4}) for PN. 
In order to exclude the contribution from background flare events, 
we extracted light curves for the full field of view \refcite{ehl-03}.
Table \ref{table1} shows the journal of our observations with {\em XMM-Newton}. 
It is known that a high-energy band is more sensitive to flare events than a
soft-energy band (Ref. \refcite{ehl-03}). We thus choose 10-12 keV energy band for MOS and 12-14 keV for PN. 
The extracted light curves were clipped to clean the contamination by
soft proton flares as decribed by Ref. \refcite{hud-05}. These Good Time
Intervals ({\scriptsize GTI}) were applied to the event lists and filtered events files are
produced. Figure \ref{fig:A2255_with_sources} shows {\em XMM-Newton}
0.3-10.0 keV energy band merged image of A2255. 
The image is raw, non-background subtracted and 
adaptively smoothed to enhance cluster emission above the background emission. 

\subsection[]{Optical Data}
Optical observations of Abell 2255 were performed with 150 cm Russian-Turkish Telescope ({\em RTT-150}). 
The telescope is located at Bakirlitepe Mountain, Antalya, south of Turkey. 
The source was observed using the ANDOR photometer and {\em TFOSC} spectrometer.
The raw data were reduced using standard procedures as described by Ref. \refcite{asl-01}.

\begin{figure*}[h]
  \begin{center}
  \includegraphics[width=11cm]{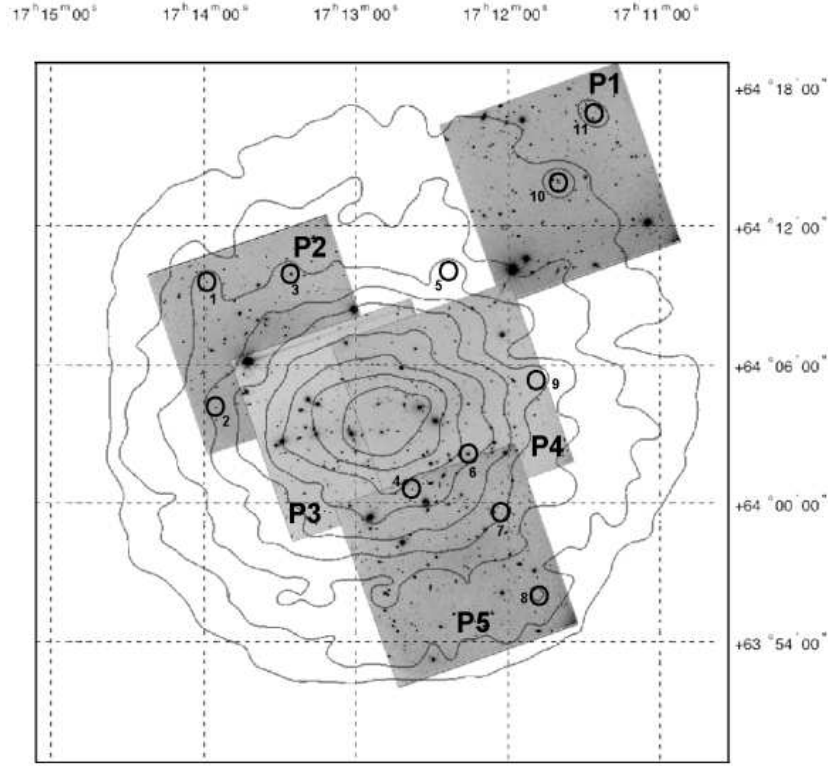}
  \end{center}
  \caption{A2255 optical {\em RTT-150} image is overlaid with X-ray
    contours from the adaptively smoothed 0.3-10 keV {\em XMM-Newton} merged image.
    Each of 5-pointing has 9$^{\prime}.1\times$9$^{\prime}$.1 square area.  
    The circles show the detected source locations.
    The image is oriented with north up and east to the left.}
  \label{fig:superpose}
  \end{figure*} 

\subsubsection[]{ANDOR Photometer}
The photometric data were taken with ANDOR CCD on 2005 June 30 and 2007 July 19.
The CCD size is 2048 $\times$ 2048 pixels at 0.24 arcsec pixel$^{-1}$ resolution.
The intended area was covered with 5 pointings.
The exposure time was 1500 sec for each frame in using Johnson B, R filters. 
The instrumental magnitudes were found by DAOPHOT\footnote{Dominion Astrophysical Observatory Photometry package} 
(Ref. \refcite{ste-87}) aperture photometry tasks in 
IRAF\footnote{IRAF is distributed by the National Optical Astronomy Observatories 
  (http://iraf.noao.edu/), which are operated by the Association of
  Universities for Research in Astronomy, Inc., under cooperative agreement
  with the National Science Foundation.}. 
These instrumental magnitudes were then calibrated using field standard stars
in the USNO A2.0\footnote{United States Naval Observatory Astrometric
  Standards} catalog. 
Figure \ref{fig:superpose} shows the superposition of 0.3-10 keV X-ray contours on 5-pointings of ANDOR. 
The contours are logaritmically spaced after the image is adaptively smoothed to enhance ICM emission.
Two bright galaxies, associated to the previous merger events, are seen at the west of the central contour.
The central contour is distorted from symetry as also reported by Ref. \refcite{sak-06}, 
which is probably caused by these two galaxies. 
 
\begin{table}[h]
\centering
\tbl{Log of TUBITAK optical telescope {\em RTT-150} observations.}
{\begin{tabular}{@{}ccccc@{}} 
\toprule
CCD   & Date         & Pointing & $\alpha$ (2000)  & $\delta$ (2000)  \\
\colrule
ANDOR & 2005.06.30  & P1    & 17:11:40.23      & 64:12:26.32     \\ 
      & 2007.07.19  & P2    & 17:13:40.99      & 64:06:04.40\\
      & 2005.06.30  & P3    & 17:13:08.90      & 64:02:35.99\\
      & 2005.06.30  & P4    & 17:12:22.00      & 64:04:45.00\\ 
      & 2005.06.30  & P5    & 17:12:25.26      & 63:56:08.67\\
\colrule
{\em TFOSC} & Grism  & Date       & Wavelength ($\AA$) &  Source \\
            & No 15  & 2006.08.23 & 3230-9120          & 2, 3, 5\\
            &        & 2007.07.19 &                    & 6\\
            &        & 2007.07.20 &                    & 4\\  
\botrule
\end{tabular} \label{table2}}
\end{table}

\subsubsection[]{{\em TFOSC} Spectrometer}
The optical spectrometric data are taken by {\em TFOSC}.
Its optical design is similar to the FOSC\footnote{http://www.astro.ku.dk/~per/fosc/index.html} series.
For this study the grism $\#$ 15 is used, which provides the maximal light efficiency 
and the widest spectral range (3230-9120 $\AA$).
Obtained spectral resolution is $\sim$ 8$\AA$.
We acquired 900 s for each source.
Wavelength calibration was performed using Neon lamp.
Reduction of spectroscopic data was done with IRAF.

\section[]{\label{analysis} ANALYSIS and RESULTS}

\subsection{\label{ICM}Temperature distribution of ICM}
First, we examined hot plasma that surrounds the cluster galaxies.
Global X-ray spectrum was extracted from a circle of radius 5 arcmin central region.
Background estimate was obtained from blank-sky event lists
(Ref. \refcite{lum-02}) after applying the same screening and filtering. 
The required response and auxiliary files for spectra were produced by rmfgen-1.53.5 and arfgen-1.66.4, respectively. 

\begin{figure*}[h]
\begin{center}
    \includegraphics[width=6.2cm]{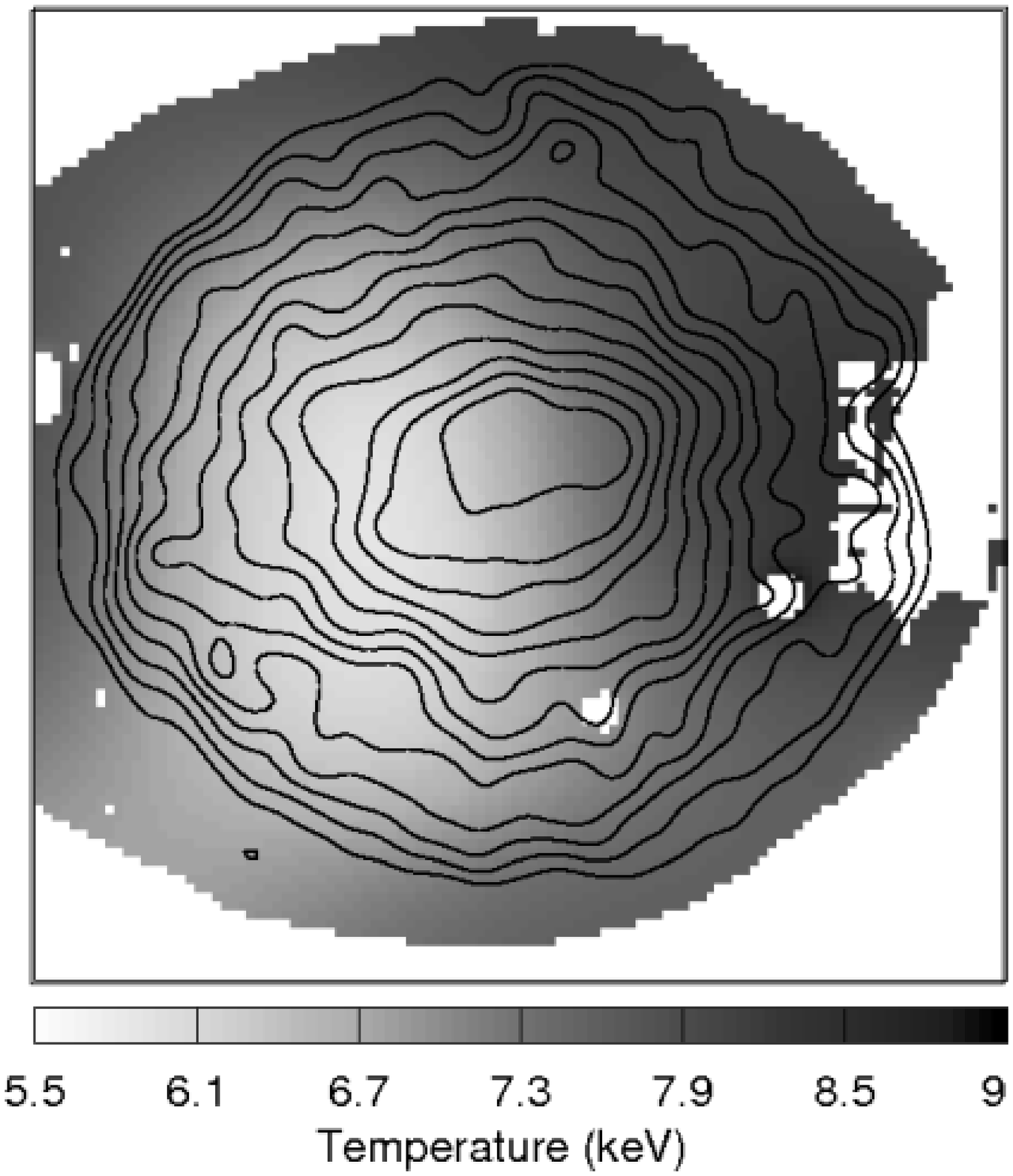}
    \includegraphics[width=6.3cm]{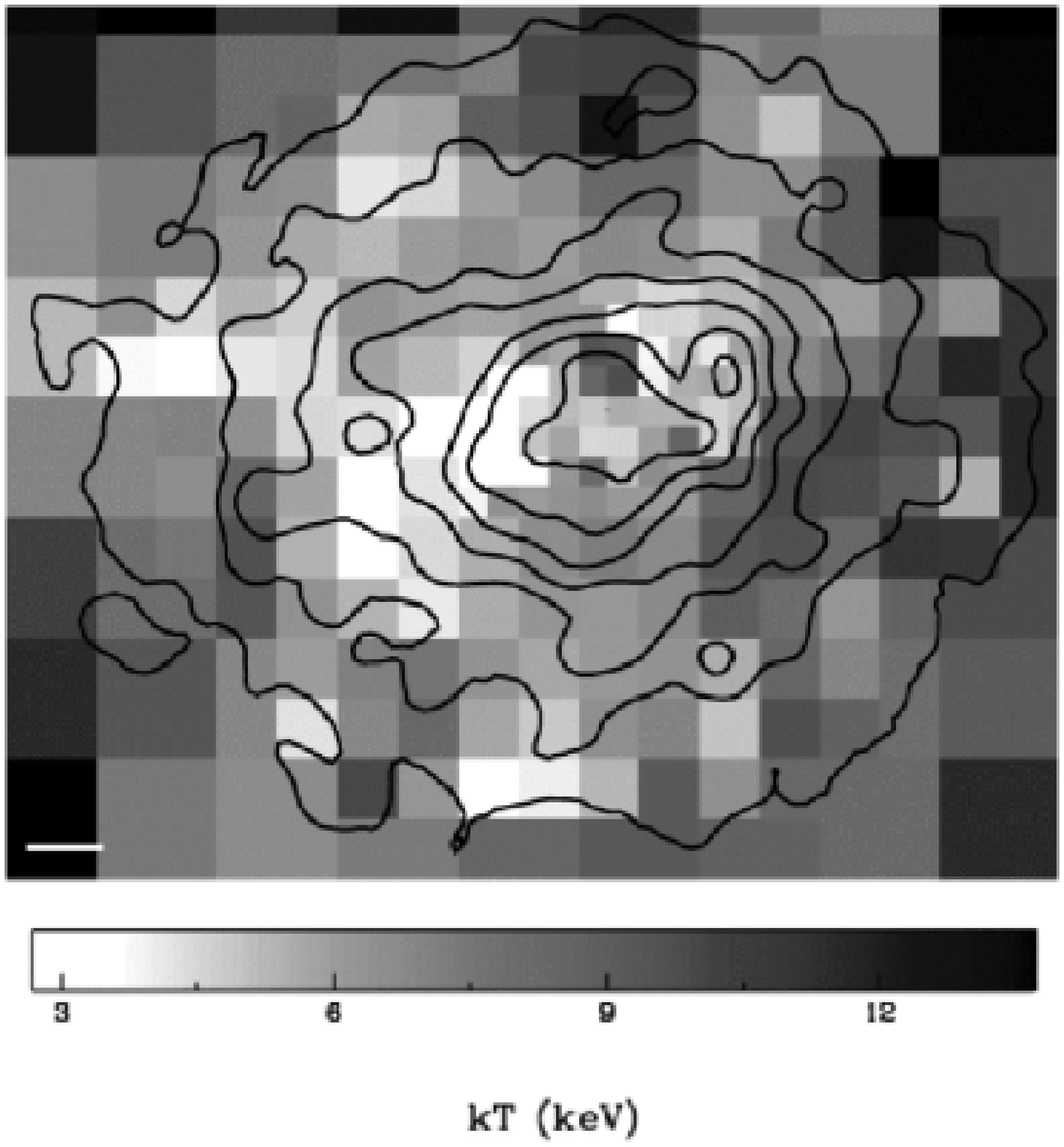}
\end{center} 
\caption{Temperature map of A2255. Color-coded scale shows temperature values. 
The east is cooler ($\sim$ 5.5 keV) than the west of ICM ($\sim$ 8.5 keV).}
\label{temperature_map}
\end{figure*}

We fitted the spectrum with a singel MEKAL model. 
It is an emission spectrum from hot diffuse gas based on the model
calculations of Mewe and Kaastra with Fe L calculations by Liedahl
(Ref. \refcite{kaa-92}).
The absorption is fixed to the Galactic value ($N_{\rm H}=$ 2.6 $\times$
10$^{20}$ cm$^{-2}$ (Ref. \refcite{dic-90}). 
The temperature value is $kT=6.7\pm0.3$ keV and metal abundance value is $Z=0.27\pm0.06$ solar. 
The temperature map is obtained by the multi-scale spectro-imaging algorithm
described in detail by Ref. \refcite{bou-04}. Figure \ref{temperature_map}
shows both our temperature map and previous map from Ref. \refcite{sak-06}.
Both maps show a nonuniform distribution of temperature. 
The best fitted temperature values are $kT=5.4\pm0.7$ keV for east and
$kT=8.4\pm0.3$ keV for west, respectively.
This result verifies the temperature variation from east to west is significant.
Our obtained global value and temperature distribution of A2255 are in good
agreement with previous work of Ref. \refcite{sak-06}. 
In this study, A2255 is selected to investigate member galaxies; therefore, a
preliminary check of A2255 ICM is enough for our objectives. 
For a detailed study of the A2255 temperature distribution and ICM dynamical state, 
the reader is referred to recent analysis results of Ref. \refcite{sak-06} and the references therein. 

\subsection{\label{det}X-Ray source detection and spectra}
Sources are detected using two different the SAS routines 
({\scriptsize  EMLDETECT} and {\scriptsize EWAVELET}) for comparison.  
The detection probability was set at $P\sim3.2 \times 10^{-5}$ by selecting
maximum-likelihood of ML=10 and 4$\sigma$ gaussian of the signal-to-noise ratio, respectively.
The detection is applied to soft (0.5-2 keV) and hard (2-10 keV) X-ray images.
The energy cuts are designed to diagnose possible intrinsic absorption and/or hard excess emission features. 
This selection also breaks the counts equally, which is statistically favorable. 
The soft and hard source-lists are merged by SAS command {\scriptsize SRCMATCH}.
A total of 11 X-ray sources are qualified in the final list from the cluster field. 
The source locations are displayed in Figure \ref{fig:superpose}.

The sources are expected to have different spectral types, and hence not all
sources were detected significantly in both soft and hard energy bands.
Of these, 7 sources were detected only in the soft band.
The other 4 sources (1, 5, 7 and 8 from Figure \ref{fig:superpose}) 
were identified in both soft and hard energy bands. 
The point-like source spectra are extracted within circles by covering 90\%
fractional energy of the energy encircled function (EEF). 
The background count rates were produced from surrounding annulus for each sources.
With this selection, we intended to remove cluster emission as well as
standard background components. Only PN data is used for the rest of the
analysis, because it has higher sensitivity than MOS cameras.

\begin{figure}
\begin{center}
    \includegraphics[width=6cm]{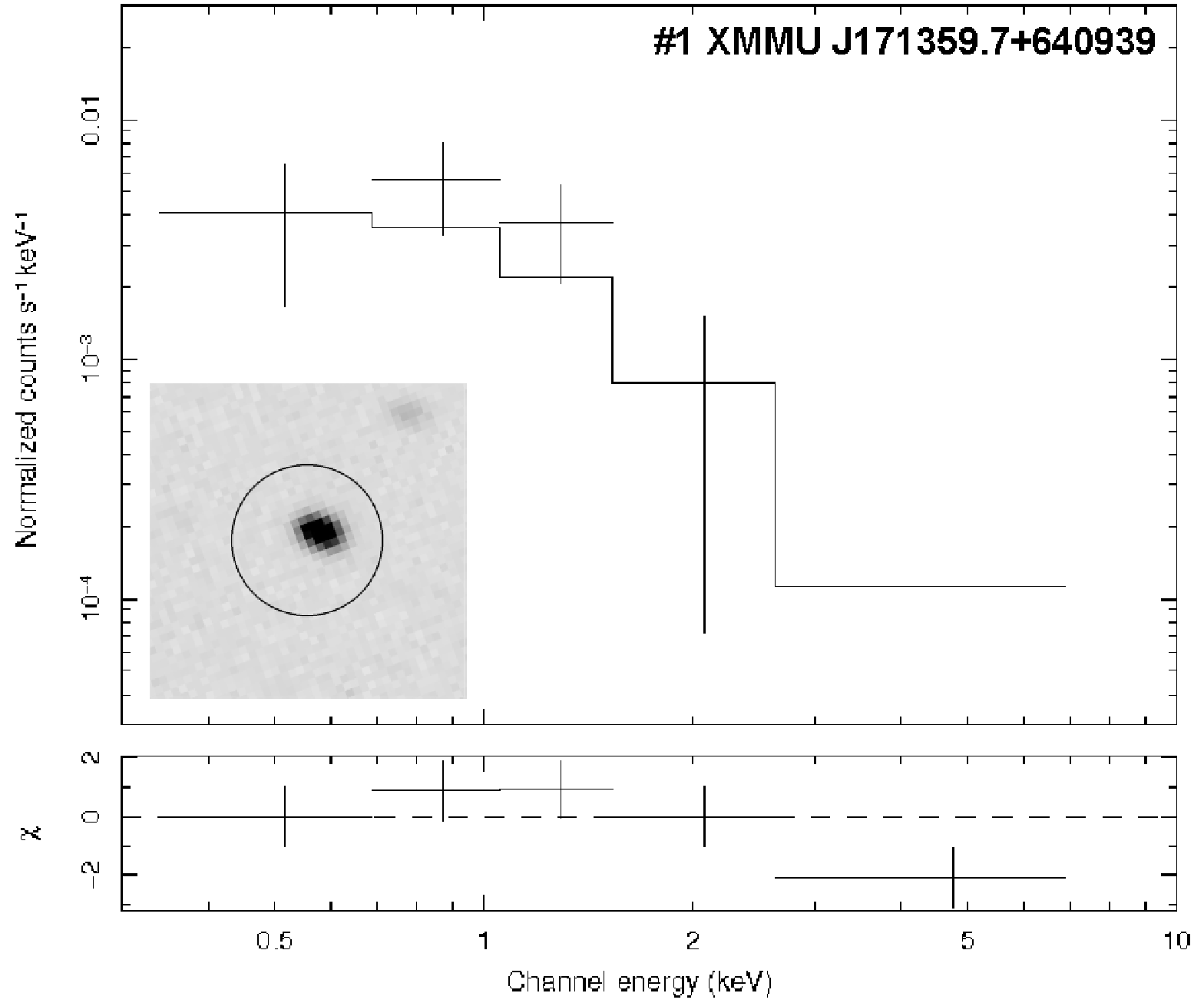}
    \includegraphics[width=6cm]{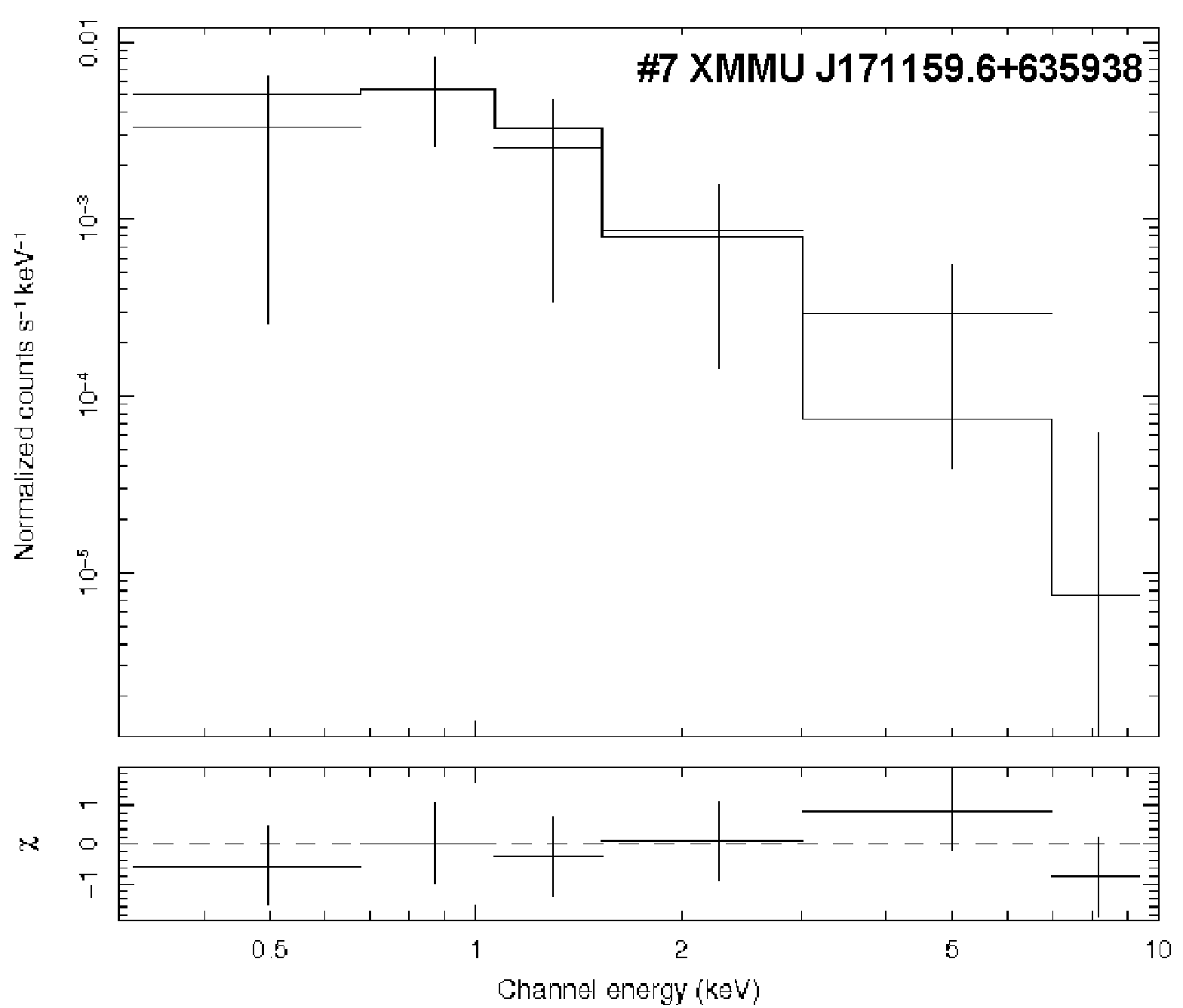}
    \includegraphics[width=6cm]{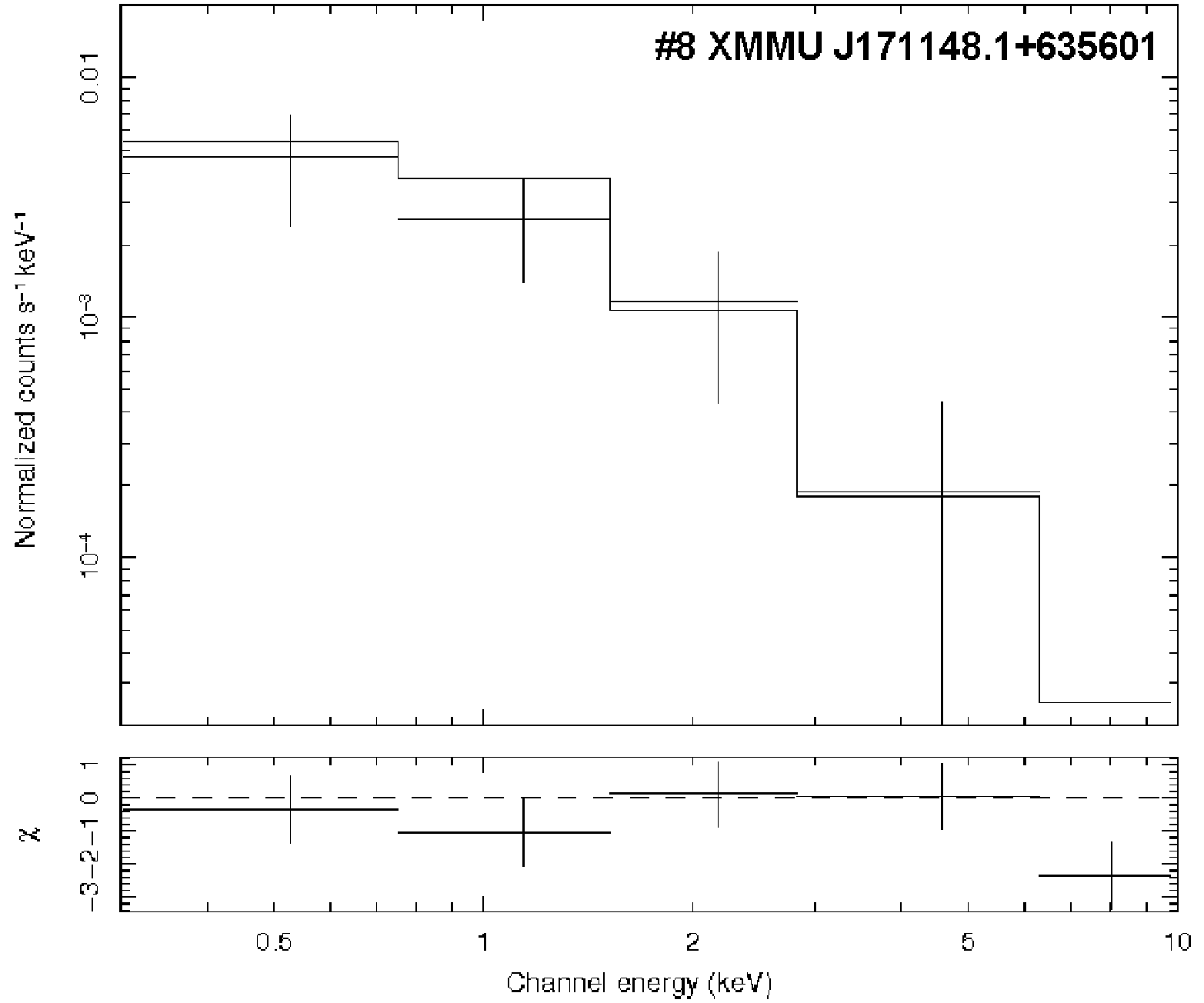}
    \includegraphics[width=6cm]{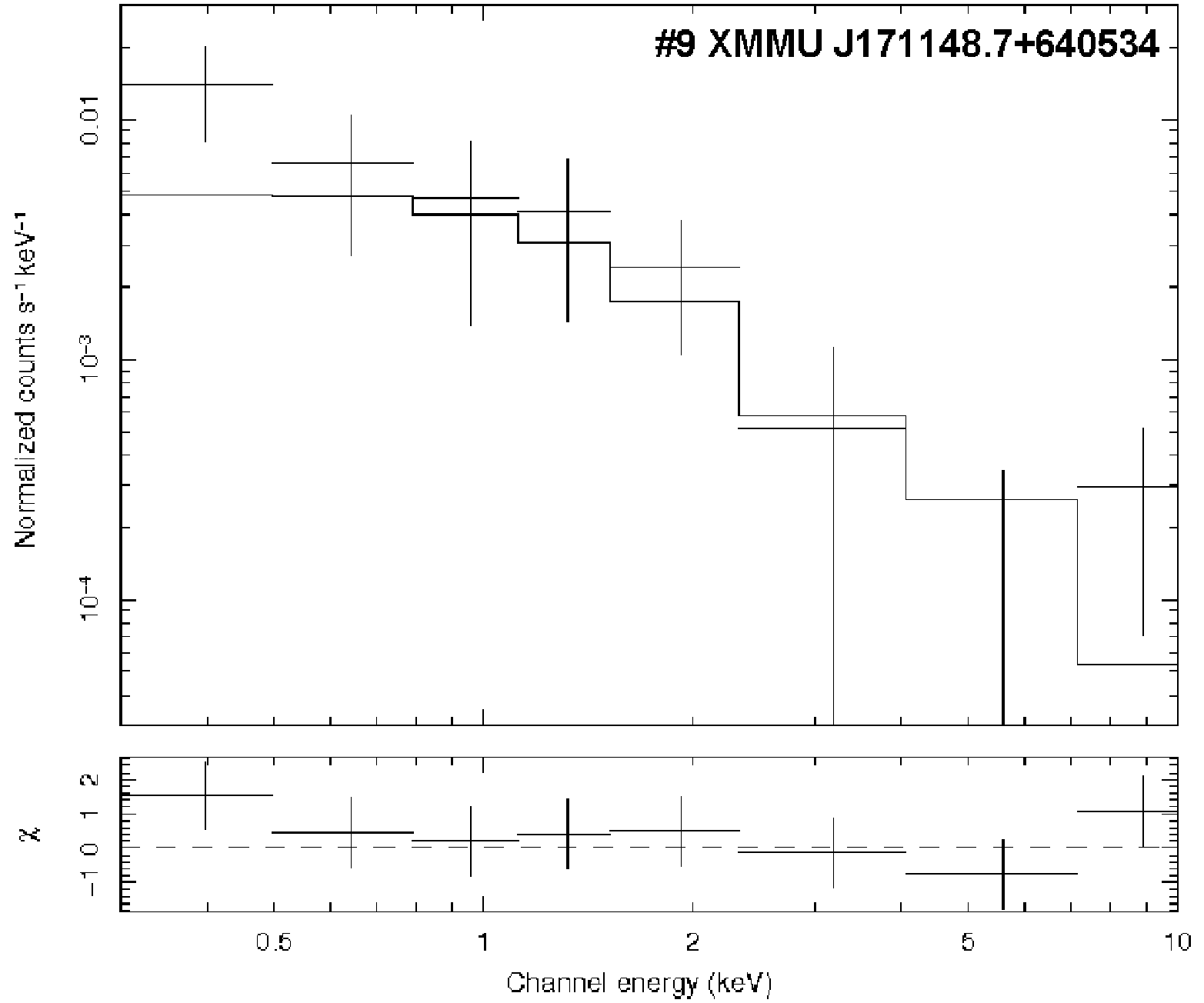}
    \includegraphics[width=6cm]{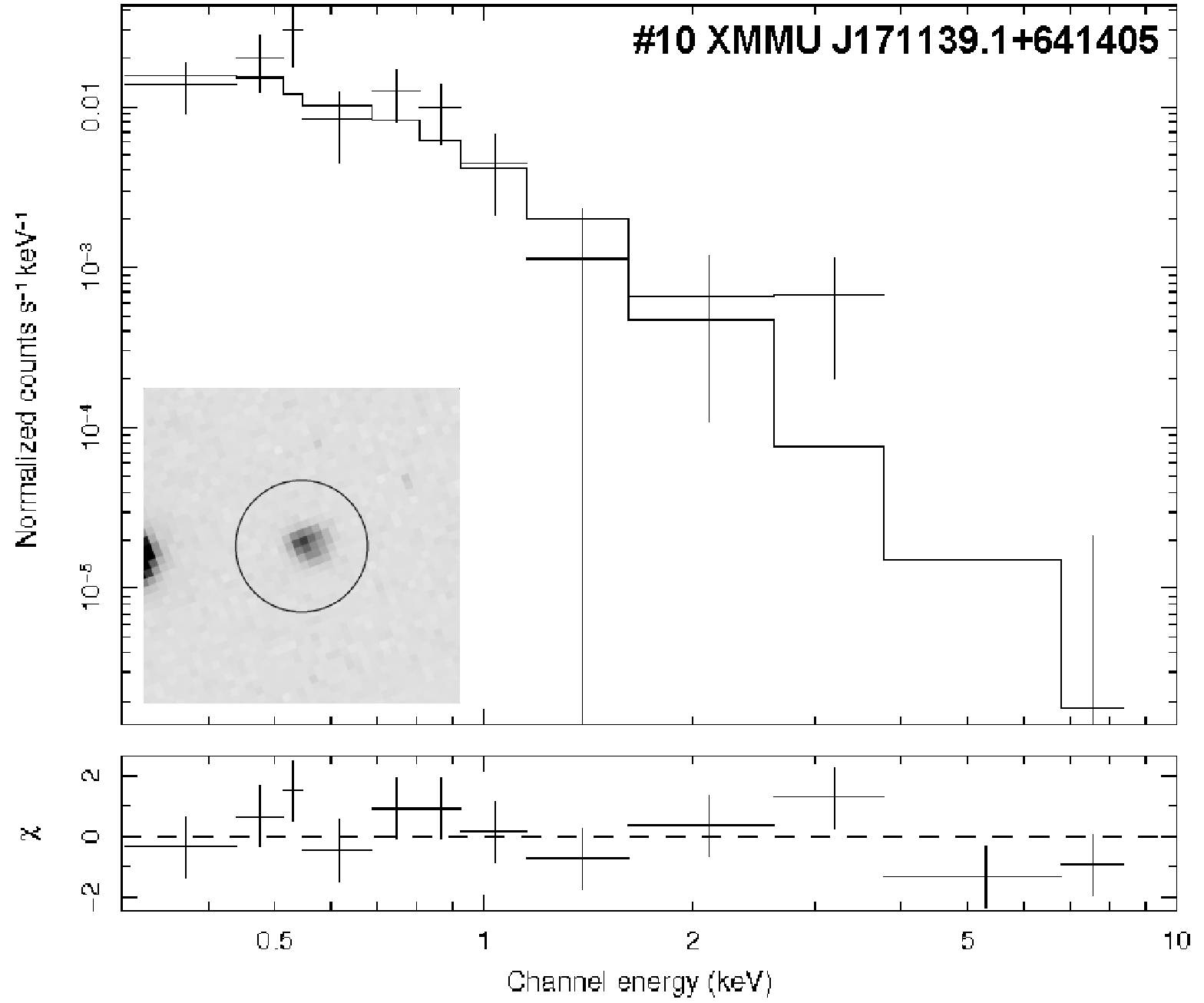}
    \includegraphics[width=6cm]{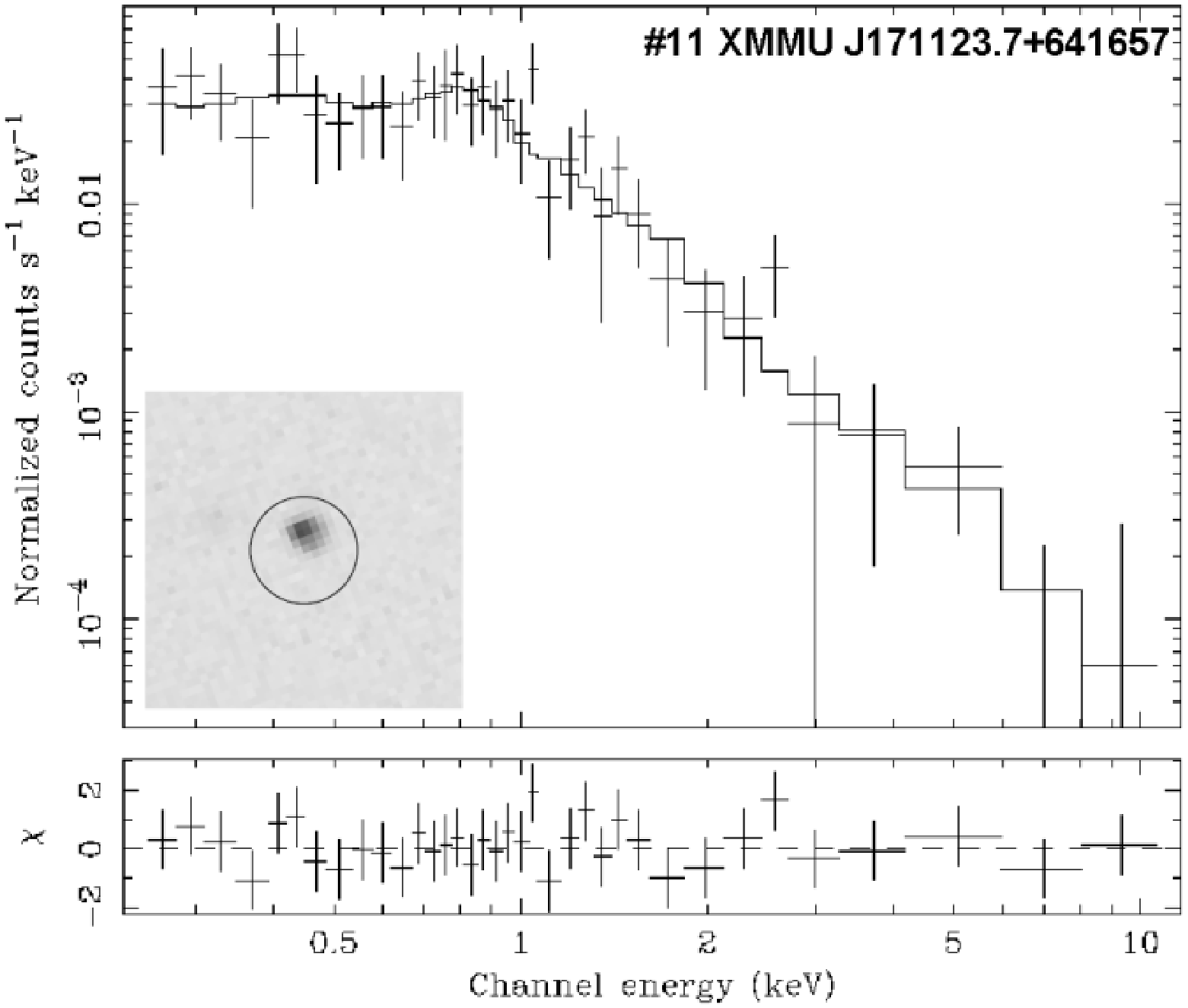}
\end{center} 
\caption{X-ray spectra for the sources (1, 7, 8, 9, 10 and 11) with no optical
  spectrum. The crosses are the data points, and the solid line is the
  best-fit power low (plus a thermal plasma for
  \#11 XMMU J171123.7+641657) model. Left-bottom images are optical
  follow-ups. The sources 7, 8 and 9 have no optical counterparts.}
\label{source_spectrum}
\end{figure}

Figure \ref{source_spectrum} shows the spectra for the sources with no X-ray
spectral information in the literature (sources 1, 7, 8, 9, 10 and 11). 
Optical followups (if there is any) of the sources are shown at the bottom left corner. 
The X-ray spectra were fitted with a single power-law emission model except
source 11-XMMU J171123.7+641657. 
There was strong deviation in the soft X-ray band (E$\le$2 keV) for the
brightest X-ray source 11. 
Single power-law model gave an index $\Gamma=$1.98$^{+0.24}_{-0.18}$ with 
${\chi}_{\nu}^{2}=$0.78 ($\nu=$31).                
Additional thermal (MEKA) model of hot gas with $kT=$0.65$^{+0.42}_{-0.26}$ keV
temperature to the power-law model ($\Gamma=$1.89$^{+0.30}_{-0.26}$) improved the
spectrum significantly (${\chi}_{\nu}^{2}=$0.59, $\nu=$32).  

The X-ray spectra for all the sources were fitted by single power-law model.
The ${\chi}_{\nu}^{2}$ values were found acceptable, corresponding to the
probability of $\ge$0.90. 
The power-law indices of the best-fit values were in the physically tolerable
range (1$<{\Gamma}<$3.5), clustering around the $\Gamma=$2.08. 
The sources 3 and 4 have too few counts ($\sim$ 110) for a spectral fit. 
For these sources, the luminosities were calculated using a power-law index
fixed to 2.0 in the spectra. The detected source properties with the best fit spectral
parameters are summarized in table \ref{table3}. The columns show (1) source
id, (2) acronmy, (3) positional error, (4) power-law index (5) F$_{X}$ and (6)
log$L_X$ at 2-10 keV, (7) reduced ${\chi}^{2}$ and (8) net counts after
background subtraction.

\begin{table}[h]
\centering
\tbl{The list of detected X-ray point sources.}
{\begin{tabular}{@{}cccccccc@{}} 
\toprule
ID & ACRONYM  & R$\sigma$ &  $\Gamma^a$   & F$_{X}$(1E-13) & log$L_X$ &
${\chi}_{\nu}^{2}$ &  Net \\
  &  &    arcsec &               & erg cm$^{-2}$s$^{-1}$ & erg s$^{-1}$   & & Count\\
(1)& (2)   &  (3)         &  (4)     &  (5)              &  (6)  & (7) & (8) \\
\colrule
 1 & XMMU J171359.7+640939 & 0.82 &  2.18$^{+0.49}_{-0.41}$ & 4.32 & 44.25 & 0.73 & 131 \\
 2 & XMMU J171342.3+640454 & 1.35 &  1.23$^{+3.53}_{-1.05}$ & 3.26 & 41.69 & 0.70 & 156 \\
 3 & XMMU J171325.4+641000 & 1.36 & 2.00$^b$                & 0.19 & 40.48 & 0.67 & 115 \\
 4 & XMMU J171236.2+640035 & 1.06 & 2.00$^b$                & 0.90 & 41.15 & 0.66 & 106 \\
 5 & XMMU J171220.9+641007 & 0.62 & 1.32$^{+0.53}_{-0.54}$  & 5.32 & ...  & 0.50 & 232 \\
 6 & XMMU J171216.2+640210 & 1.41 & 1.72$^{+1.24}_{-0.98}$  & 5.56 & 41.94 & 0.75 & 289 \\ 
 7 & XMMU J171159.6+635938 & 2.17 & 3.51$^{+2.36}_{-2.44}$  & 0.68 & 41.08 & 0.63 & 190 \\
 8 & XMMU J171148.1+635601 & 0.81 & 1.79$^{+1.23}_{-0.72}$  & 1.85 & 41.49 & 0.82 & 127 \\
 9 & XMMU J171148.7+640534 & 0.80 & 1.98$^{+1.17}_{-0.86}$  & 1.68 & 41.45 & 0.52 & 232 \\
10 & XMMU J171139.1+641405 & 2.25 & 3.09$^{+0.81}_{-0.63}$  & 0.76 & 43.17 & 0.67 & 181 \\
11 & XMMU J171123.7+641657 & 0.56 & 1.89$^{+0.30}_{-0.26}$  & 15.52 & 42.39 & 0.59 & 372 \\
\botrule
\end{tabular} \label{table3}}
\end{table}

\subsection{\label{log}log($N$)-log($S$)}
\label{counts}

\begin{figure}[h]
  \begin{center}
\includegraphics[width=7cm,angle=-90]{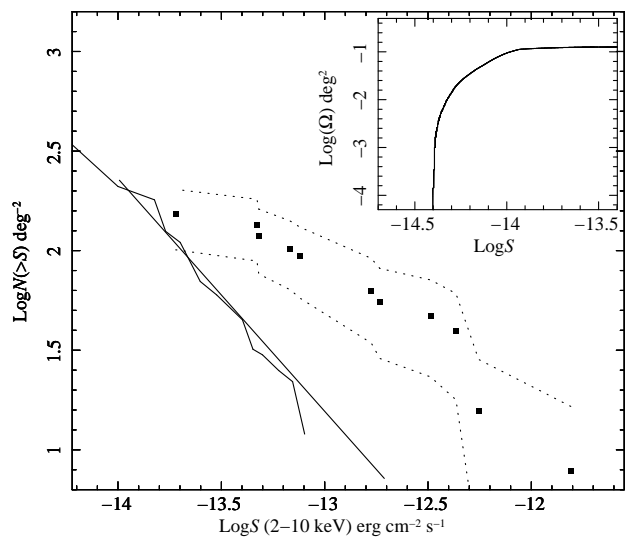}
   \end{center}
  \caption{log($N$)-log($S$) of the X-ray selected sources from A2255 field (filled-boxes)
    compared with the field of Lockman Hole (solid broken line) by
    Ref. refcite{has-01}. 
    The line shows the power-law model (Ref. refcite{hud-06}).
    The dotted lines show 1$\sigma$ confidence level of statistical constrains and
    10\% calibration errors (Ref. refcite{kir-06}). The top corner shows the survey area,
    $\Omega$ deg$^2$ vs flux limit.} 
\label{fig:logN_logS}
\end{figure}

The flux values in hard band (2-10 keV) derived from spectral fitting
have provided good measurements of the source number counts, or log($N$)-log($S$)
relationship. The sample includes 11 sources with fluxes between 10$^{-11.8}$ and
10$^{-13.7}$ ergs cm$^{-2}$ s$^{-1}$. 
The results are compared with Lockman Hole (non-cluster field).
We acknowledge that the number of sources is admittedly low, but it can be
regarded as being a rational method to infer appreciable knowledge on cluster galaxies.
The cumulative source number per area (degree$^{-2}$) can be calculated as the sum of
the inverse areas of all sources brighter than flux $S$, $N(>S)$. Using the
integrated formula of source numbers in units of deg$^{-2}$,
$N(>S)=\!\!\sum_{i=1}^{n}\!\!\frac {1}{\Omega_i} ($deg$^{-2})$, 
where $n$ is the detected source number and $\Omega_i$ is the sky coverage
for the flux of the $i$-th source, the relation between source number and flux
is derived. It is a well known fact that each pixel on detector is not equally sensitive. 
The survey area decreases with the flux, and hence the detection
probability decreases as off-axis angle increases. 
To determine the number counts per area with precision, it is necessary to correct
for the incompleteness of the sample. 
Figure \ref{fig:logN_logS} shows a comparison of the obtained
log($N$)-log($S$) plot for A2255 and the Lockman Hole results
(Refs. \refcite{has-01}, ).
The dotted-lines include statistical and systematic instrumental errors. The
statistical error was represented with 1$\sigma$ poisson errors
($\sigma_{dN}=\sqrt{ \sum_{i=1}^{n} (1/{\Omega_i})^2 } $deg$^{-2}$) in the source numbers.
The systematic instrumental errors related to vignetting and the point spread
function (PSF) is considered as 3.5\% by Ref. \refcite{sax-03} for an off-axis
source. To be conservative, we favor to use 10\% absolute calibrational error
defined by Ref. \refcite{kir-06}. 
The source numbers at each flux level was corrected by considering the corresponding sky coverage.
Each flux value in the plot was corrected by using the sensitivity map (top
left corner of Figure \ref{fig:logN_logS}). The map shows the survey area,
$\Omega$ deg$^2$, as a function of the limiting fluxes in the 2-10 keV
band. The counting rates were derived with the assumption of $\Gamma =$ 2.0
and the galactic value of the hydrogen absorption column density
(Ref. \refcite{dic-90}). 

Figure \ref{fig:logN_logS} also includes the Lockman Hole X-ray source study by Ref. \refcite{has-01}.  
One of the other recent works on Lockman Hole by Ref. \refcite{hud-06} also
investigated the source properties by modelling. 
Based on their study, single power law assumption converts the field source numbers,  
$N (>S) = 1.45 {\times} 10^{-14} \times S^{-1.16_{-0.25}^{+0.20}}$, where the
slope is ${\alpha}=$ 1.16$_{-0.25}^{+0.20}$ and the normalization is $K= 1.45{\times} 10^{-14}$ deg$^{-2}$. 
The solid-line in Figure \ref{fig:logN_logS} shows the model-line for Lockman Hole. 

The log($N$)-log($S$) plot shows large population of X-ray sources in A2255
cluster clearly. The significance increases at the brighter end. The source
density value corresponds to 102$\pm$39 sources deg$^{-2}$ for the cluster
field around the flux value of $\sim$ $10^{-13}$ erg cm$^{-2}$ s$^{-1}$.
From the above mentioned equation, a density of 22$\pm$3 sources deg$^{-2}$ from Lockman Hole
was calculated at this flux level. The error corresponds to field-to-field variation of the
blank field, which is known as cosmic variance. Based on the most sensitive
mesurement to the date, it is reported to be less than 15\% in the 2-10 keV
band (Ref. \refcite{cappe-05}). If we perform a descriptive calculation by
using the minumum possible source density from A2255 and the maximum source
density for the Lockman Hole, the number density is 38 sources deg$^{-2}$ for A2255. 
This worse-case scenario calculation confirms that $\sim$ 40\% of the detected X-ray sources are A2255 members.
It corresponds to at least a factor of 4 times higher source number excess from
the cluster region at this flux level. 
However, it vanishes gradually at the fainter end as expected.

\subsection{\label{TUG}Optical imaging and photospectrum}

X-ray selected sources from the cluster field are observed by ground based
{\em RTT-150} telescope for 1500s exposures with five pointings.
Figure \ref{fig:superpose} shows the pointings with detected 11 x-ray sources (circles).
Of these, 7 sources were found with optical counterpart (limiting magnitude $R$=18.48). 
The locatization accuracy was high enough that optical followup sources were
not tested for positional match. The largest shift between optical and x-ray
peak was $\Delta R$=4.36 arcsec for source 4. 
This was in the confidence range considering R$\sigma$ errors. 
We assigned $B$ and $R$ magnitudes to the whole spectroscopic sample. 
Important parameters derived from optical observations are given in Table \ref{table4}. 
The columns are  (1) id; (2) acronmys;  
(3) $m_R$ apparent $R$ magnitude; 
(4) $m_B$ apparent $B$ magnitude; 
(5) $M_B$ absolute $B$ magnitude calculated from apparent magnitude and distance, as $M_B = m_B + 5 - 5 log(d)$; 
(6) $B$ luminosity in solar-units, defined as log $L_B = -0.4 (M_B - 5.41)$; 
(7) optical redshifts calculated from {\em TFOSC} spectroscopy (see \ref{notes} for detail). 

\begin{table}[h]
\centering
\tbl{Optical {\em RTT-150} observations for the X-ray selected cluster galaxies.}
{\begin{tabular}{@{}ccccccc@{}} 
\toprule
ID & ACRONYM        & $m_R$ & $m_B$ & $M_B$ & log ($L_B/L\odot$) &   $z$ \\
(1) & (2)        & (3) & (4) & (5) & (6) &   (7) \\
\colrule
1 & XMMU J171359.7+640939 & 17.66 & 17.98 & ... &  ...   & ... \\
2 & XMMU J171342.3+640454 & 15.03 & 16.64 & -20.77 & 10.47   &  0.0824 \\
3 & XMMU J171325.4+641000 & 13.92 & 15.80 & -21.62 & 10.81  & 0.0811  \\
4 & XMMU J171236.2+640035 & 17.10 & 18.68 & -18.73 & 9.66   & 0.0883 \\
6 & XMMU J171216.2+640210 & 14.85 & 17.00 & -20.41 & 10.33   & 0.0851 \\ 
 10 & XMMU J171139.1+641405 & 15.17 & 18.98 & ... & ...   & ... \\
 11 & XMMU J171123.7+641657 & 18.48 & 19.09 & -18.32 & 9.49   & ... \\
\botrule
\end{tabular} \label{table4}}
\end{table}

\begin{figure*}
  \begin{center}
  \includegraphics[width=8.7cm]{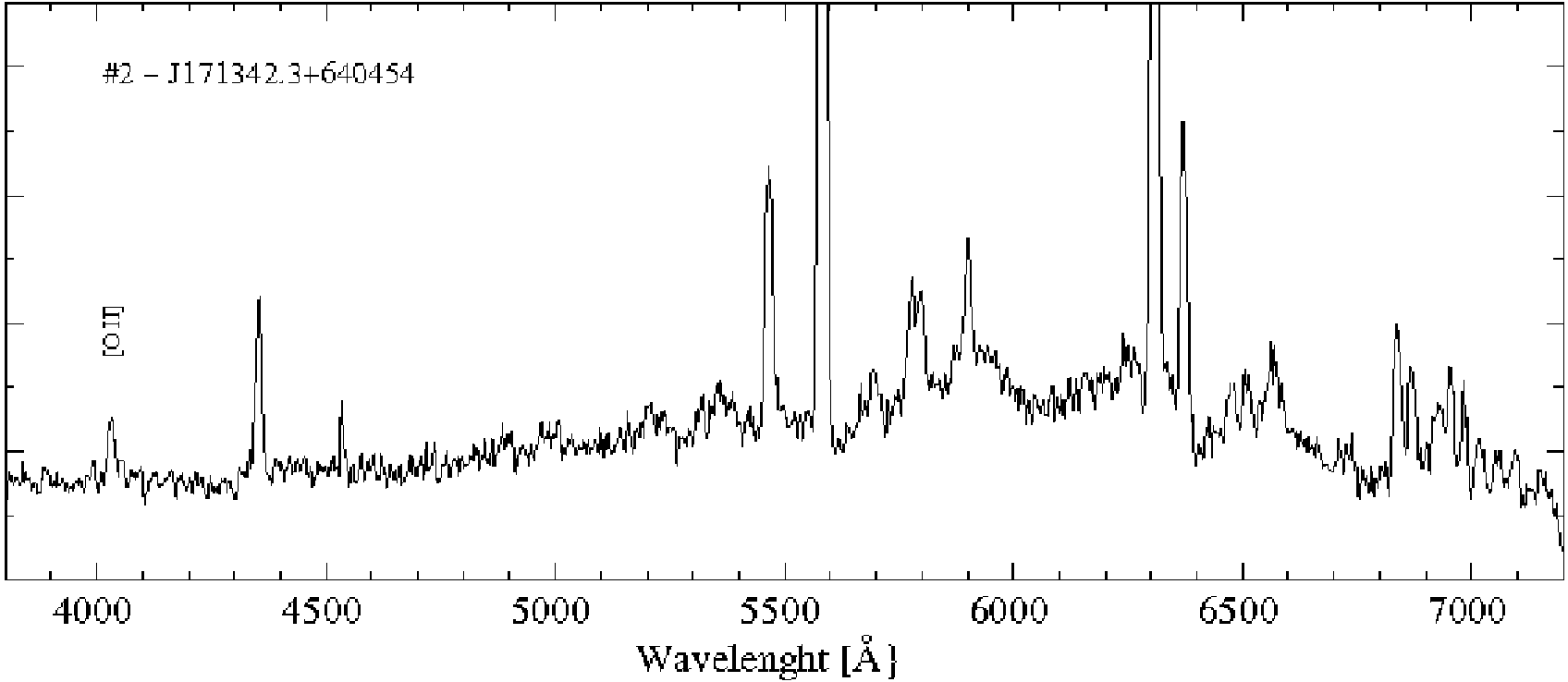}
  \includegraphics[width=3.8cm]{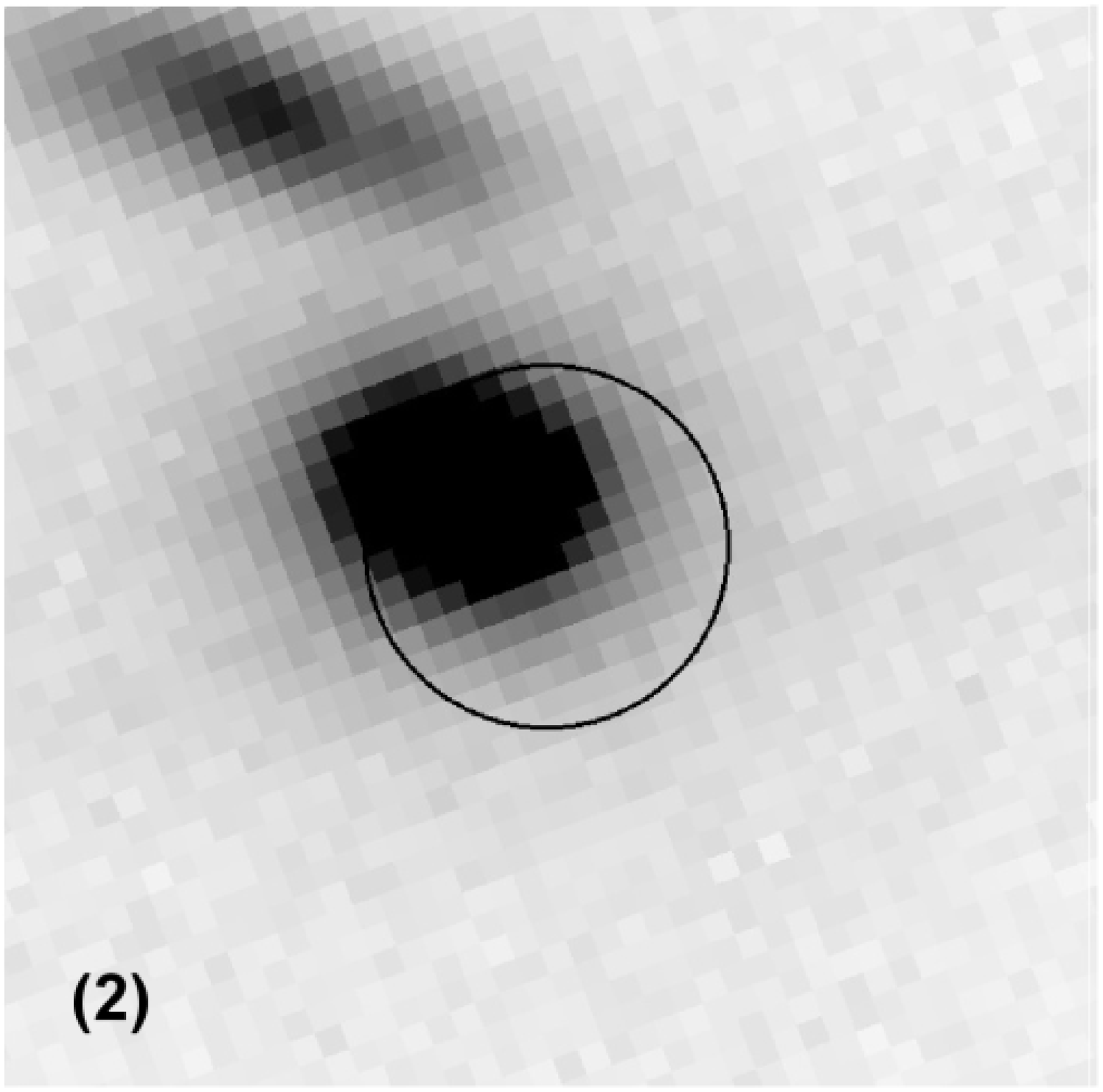}
  \includegraphics[width=8.7cm]{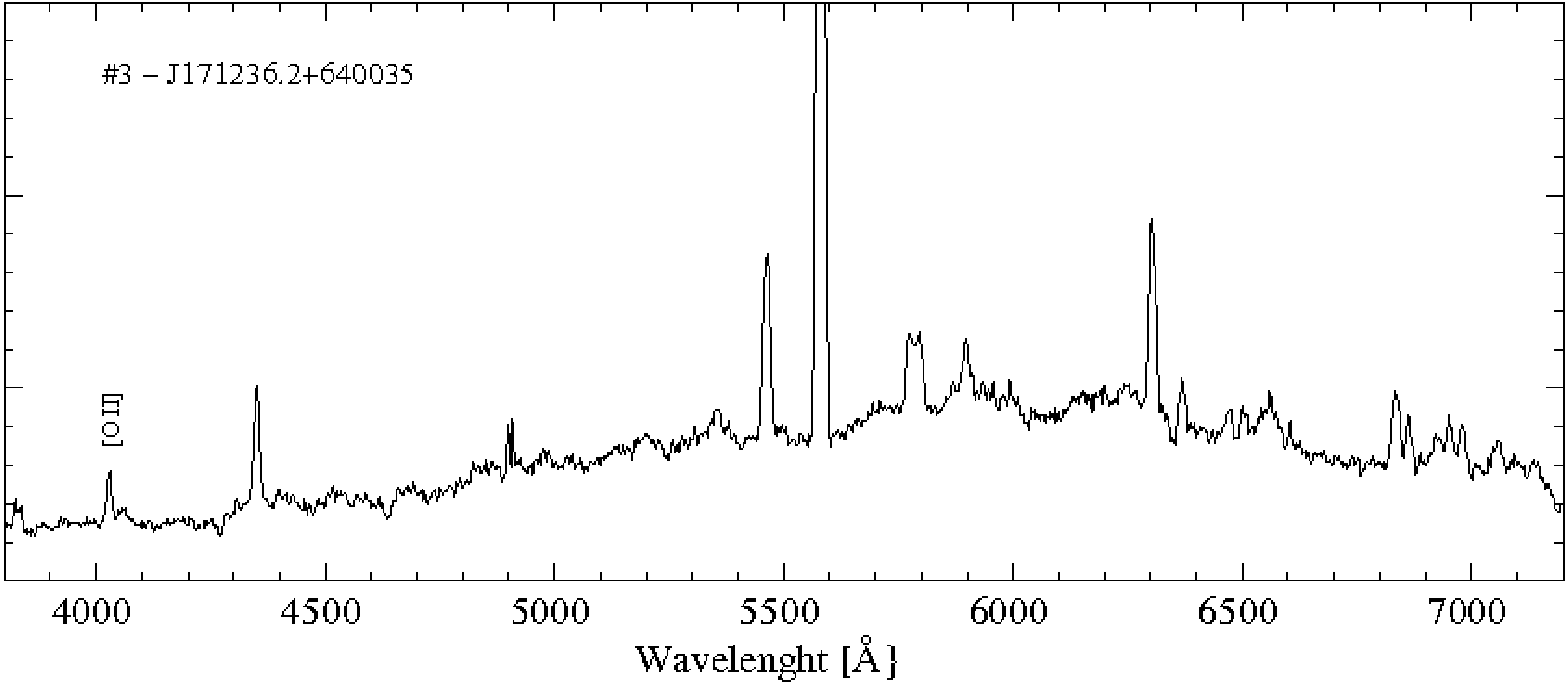}
  \includegraphics[width=3.8cm]{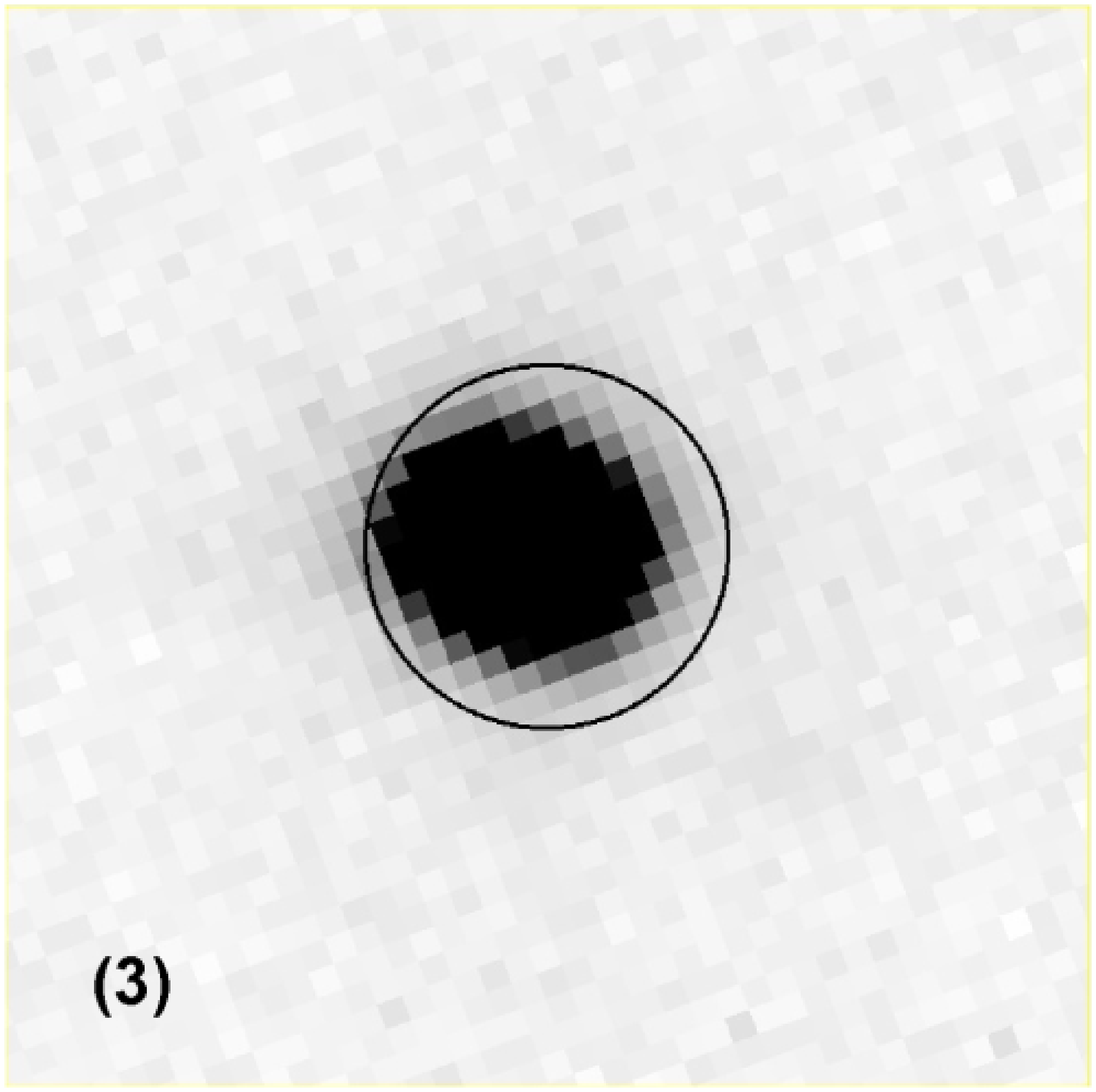}
  \includegraphics[width=8.7cm]{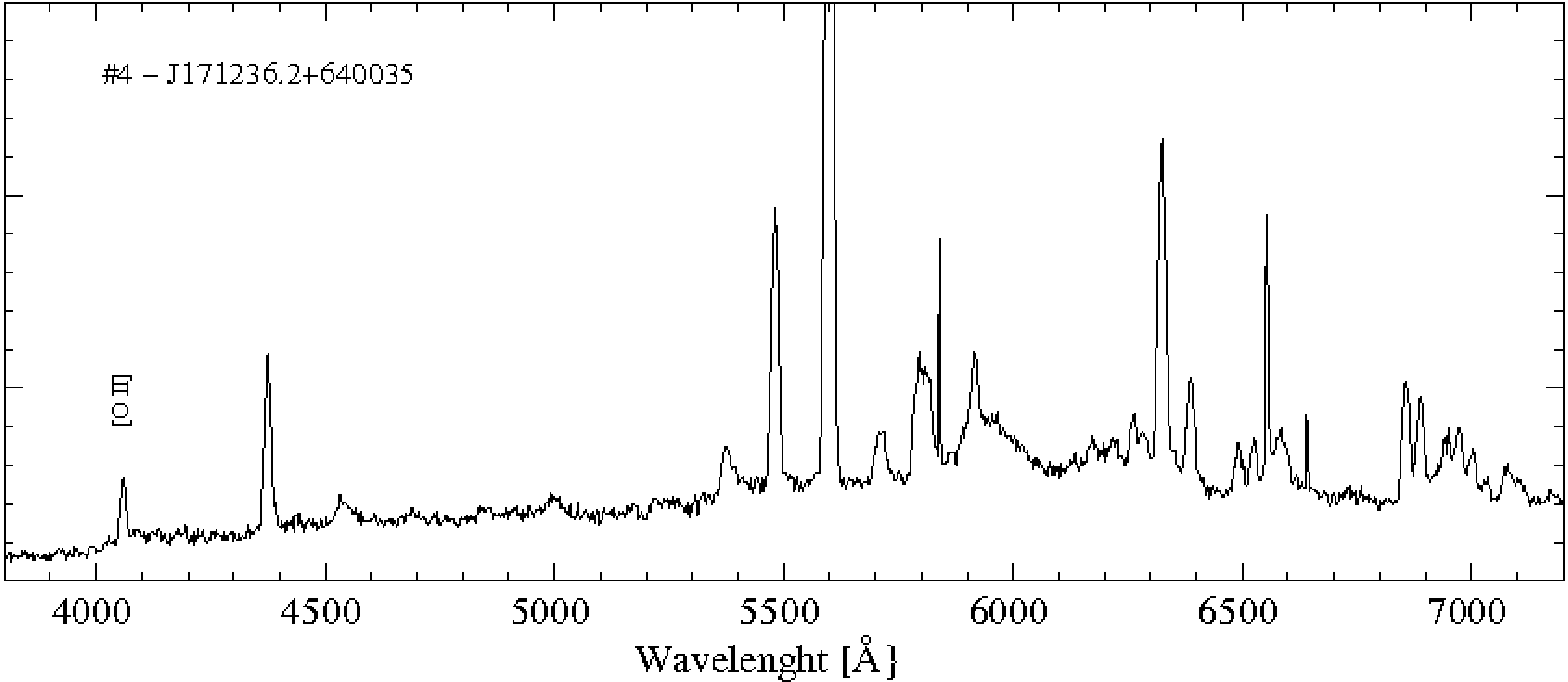}
  \includegraphics[width=3.8cm]{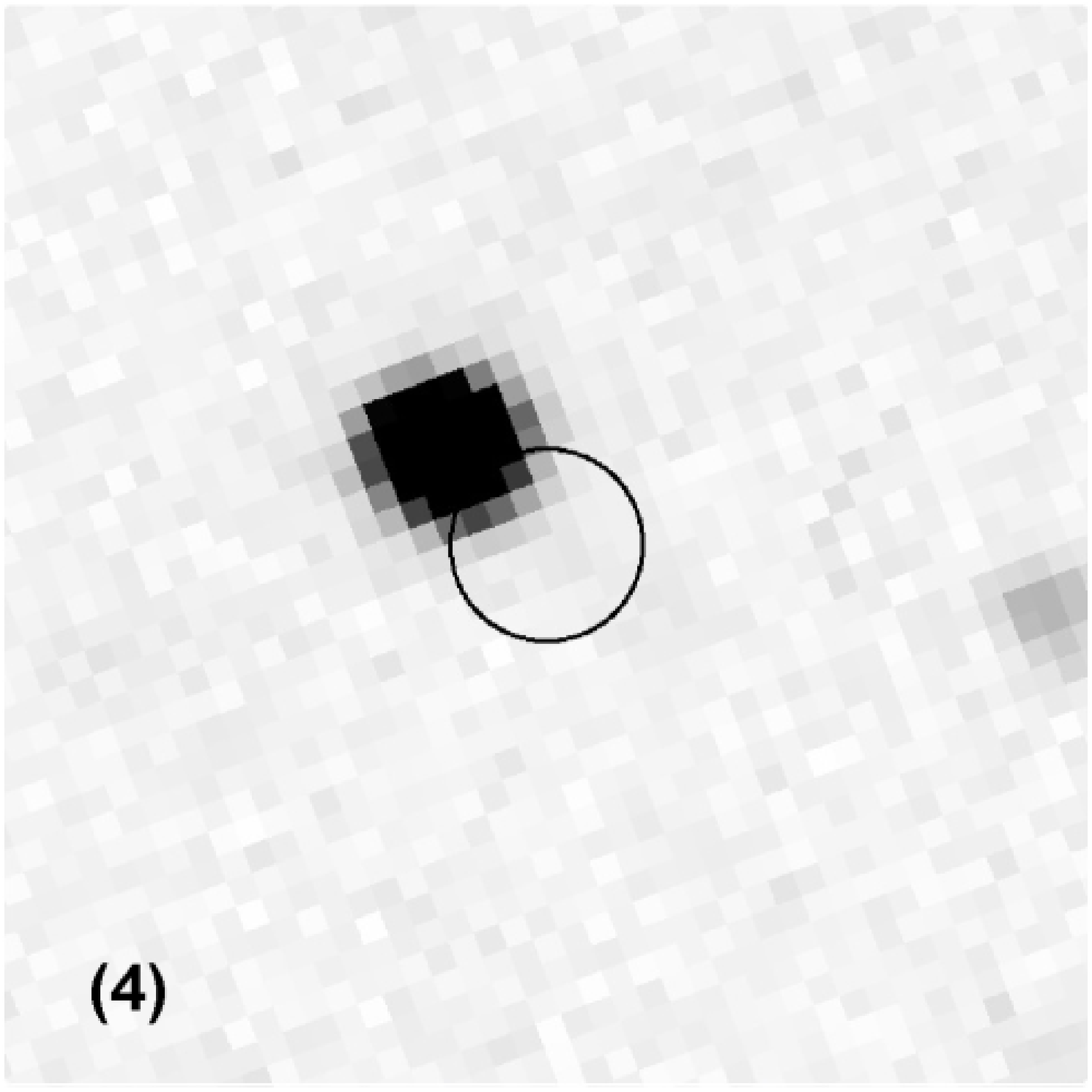}
  \includegraphics[width=8.7cm]{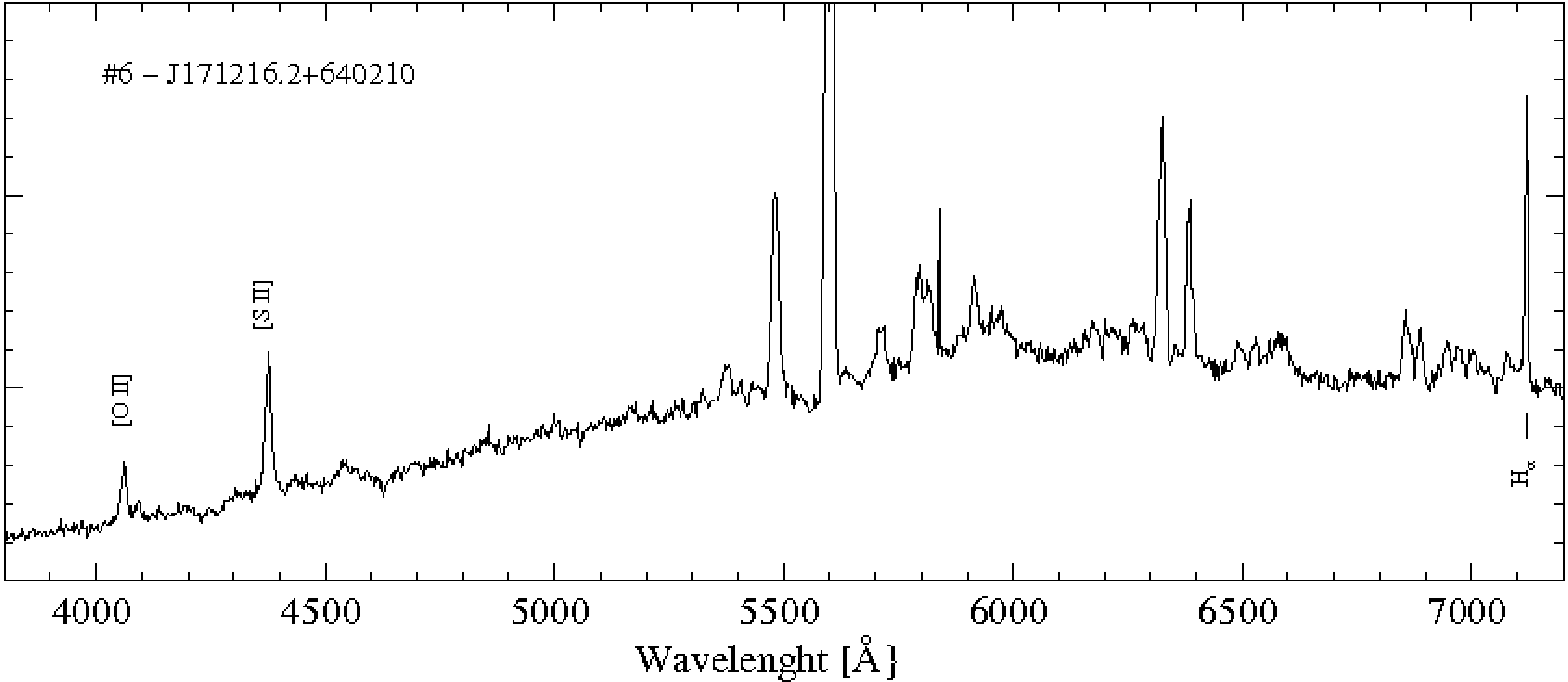}
  \includegraphics[width=3.8cm]{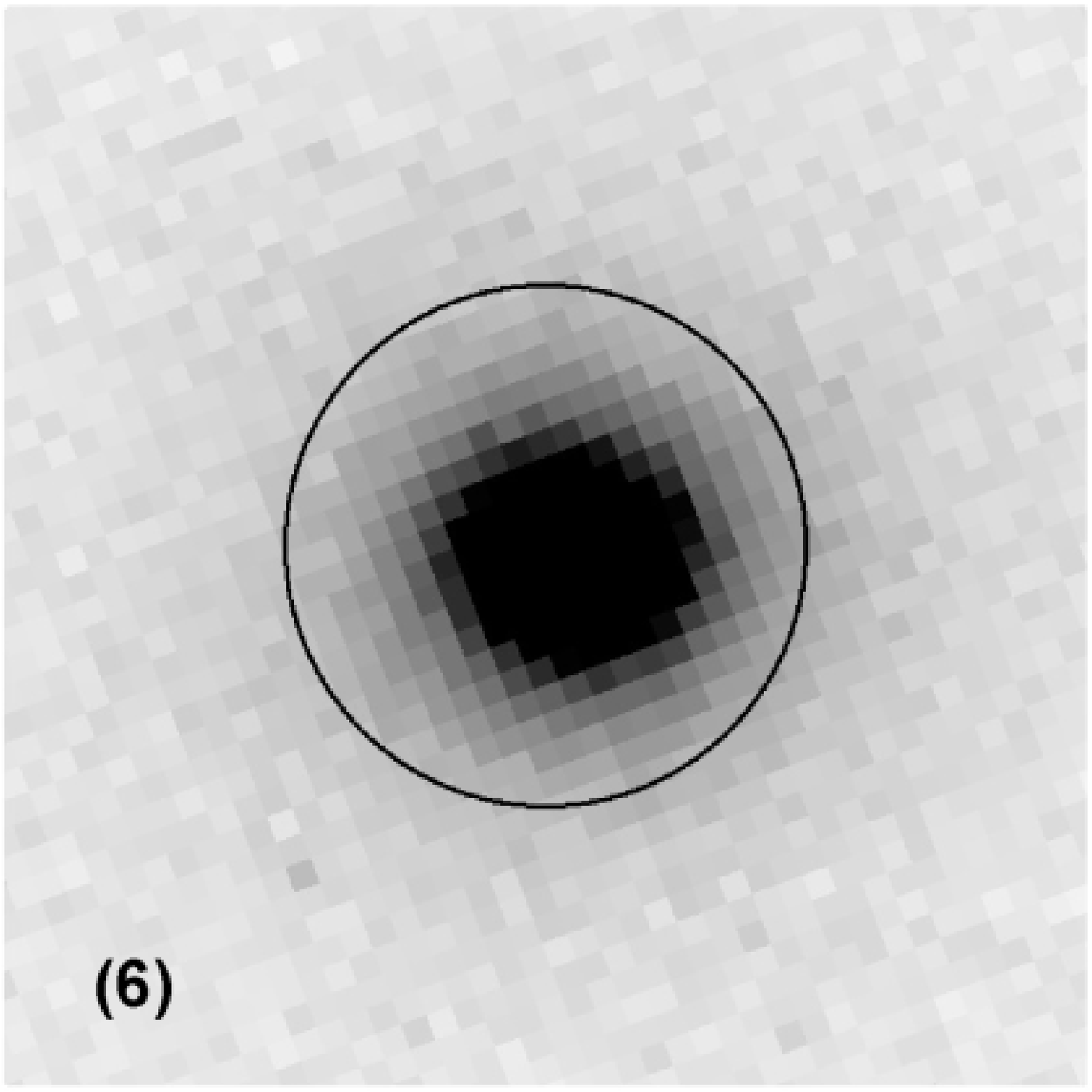}
  \end{center}
\caption
{The images of $R$-band RTT-150, 20$^{\prime\prime}$ $\times$
  20$^{\prime\prime}$ panel sized cutouts. The circles indicate the source
    extends in X-rays. It is defined as $\sigma$$\sqrt{8ln2}$, which corresponds to the FWHM for a gaussian.
    The images are oriented with north up and east to the left.}\label{fig:optic}
 \end{figure*} 

\begin{figure*}[h]
  \begin{center}
 \includegraphics[width=8.7cm]{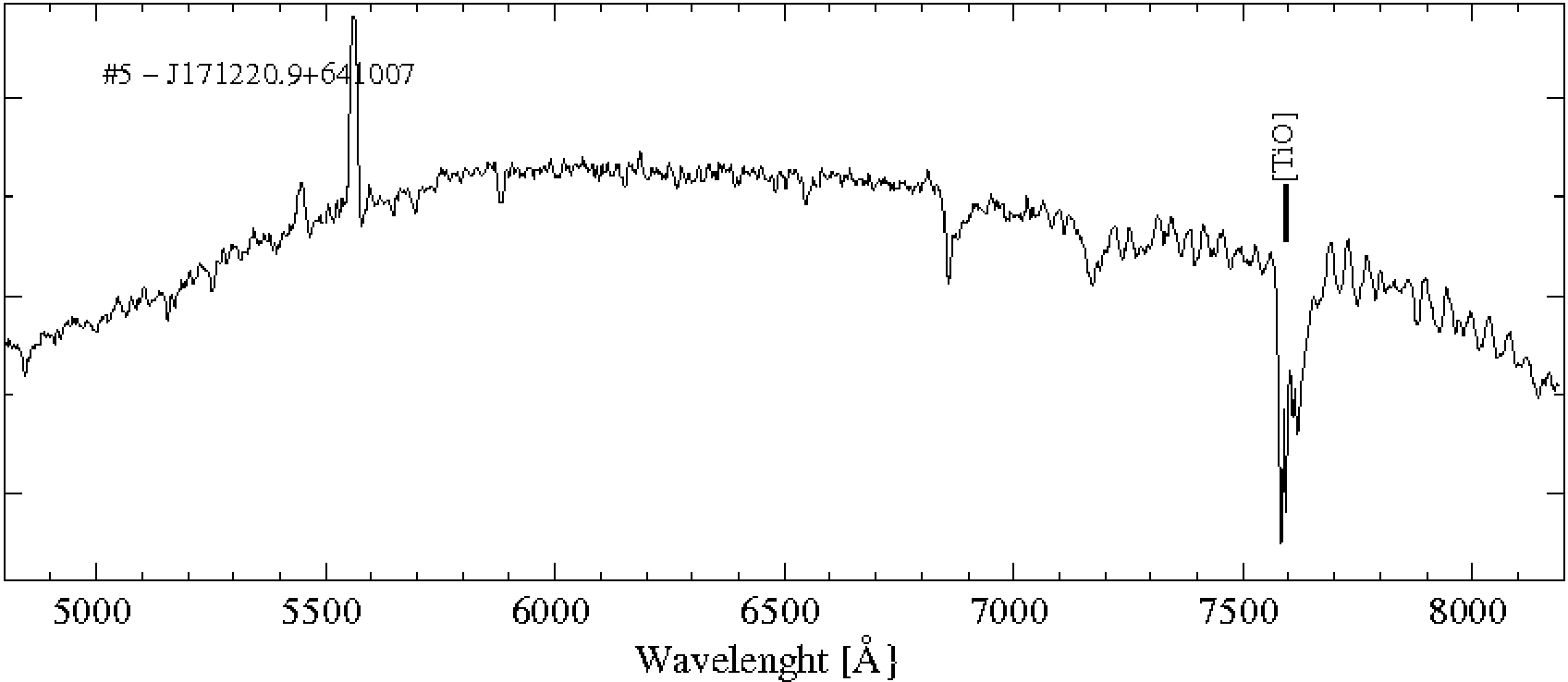}
   \end{center}
 \caption{{\em TFOSC} optical spectrum for star \#5-XMMU J171220.9+641007.}
\label{fig:star}
\end{figure*} 

Sources 1 (XMMU J171359.7+640939) and 10 (XMMU J171139.1+641405) were
identified as background objects. The reshifts are $z$ = 1.363
(Ref. \refcite{sch-02}) and $z$ = 0.925 (Ref. \refcite{ric-04}), respectively. 
Source 11 (XMMU J171123.7+641657) had the faintest optical counterpart.
Sources 2, 3, 4 and 6 were studied with {\em TFOSC} spectrometer and redshift
values were obtained.
 Figure \ref{fig:optic}
shows the optical spectra with [O $_{\rm II}$] emission lines and {\em ANDOR}
$R-$band images for the sources. The images are 20$^{\prime\prime}$
$\times$ 20$^{\prime\prime}$ sized cutouts. 
The spectral wavelength range of {\em TFOSC} Grism 15 was enough for the coverage of
redshifted [O $_{\rm II}$] $\lambda$3727, H$_{\rm \alpha}$ $\lambda$6563, and 
[S$_{\rm II}$] $\lambda$6716, $\lambda$6731 doublet. 
The spectra are severely comprised by plenty of unkown emission lines.
[O $_{\rm II}$] emission line is known as strong indicator of galaxy colour and
good tracer of star formation (Ref. \refcite{gal-89}). 
Therefore we principally applied [O $_{\rm II}$] line for the redshift
calculations where the spectra is less ambiguous (Fig. \ref{fig:optic}). 

The source 5 was not pointed in {\em ANDOR} observations (see
Fig. \ref{fig:superpose}), but {\em TFOSC} spectrum was studied for this source. 
Figure \ref{fig:star} shows the spectrum for source 5 (XMMU J171220.9+641007). 
Strong TiO absorbtion was interpreted as optical spectra of M4 star.
The absorption band appears near 7590 $\rm \AA$. 
The X-ray spectrum of the source was fairly hard ($\Gamma$ $\sim$ 1.3) which
is caused by magnetic activity (Ref. \refcite{bra-05}). 
It has an X-ray flux of 5.32 $\times$ 10$^{-13}$ ergs cm$^{-2}$ s$^{-1}$. 
Based on this result, the source was not reobserved with {\em ANDOR} photometer. 

\section{\label{discussion}Discussion}

We detected 11 X-ray sources from A2255 outskirts. 
Based on the log($N$)-log($S$) (See Figure \ref{fig:logN_logS}) 
the source number was estimated to be about 4 times higher relative to the field
at $F_X$ = 1 $\times$ $10^{-13}$ ergs cm$^{-2}$ s$^{-1}$ flux limit. The X-ray
luminosity values range between 40.48 $<$ log(L$_X$) $<$ 42.39 ergs
s$^{-1}$, clustering at log(L$_X$) = 41.37 ergs s$^{-1}$.
This excess indicates either an elavated X-ray emission or/and higher
population of point sources from the cluster field.

X-ray emission from early-type galaxies has two origins: thermal emission from
interstellar hot gas (ISM) and non-thermal energy release from accretion processes. 
If the matter accretes onto lower mass ($\le$ 1$M_{\odot}$) of neutron star or black hole the system is known as low-mass X-ray binary (LMXB). 
If the accretion is due to supermassive blackhole ($\sim$10$^8$ $M_{\odot}$) then the source is called active galactic nucleus (AGN). 
X-ray luminosity of thermal emission from hot halo is log(L$_X$) 37$\sim$38 ergs s$^{-1}$
(Ref. \refcite{bla-01}). Expected luminosity from accetion processes is
log(L$_X$) 38$\sim$39 ergs s$^{-1}$ for LMXB and log(L$_X$) $\ge$ 40 ergs
s$^{-1}$ for AGNs (Refs. \refcite{bla-01}, \refcite{ran-06}). 
Regarding the luminosity range observed in our study (log(L$_X$) $\ge$ 40.48
ergs s$^{-1}$), we believe that x-ray emission is dominated by discrete sources
(LMXBs) rather than hot halo (e.g., Refs. \refcite{leh-07}, \refcite{dav-06},
\refcite{sar-01}, \refcite{bla-01} and references therein) unless a
central AGN contributes to the emission. 

\begin{figure}
  \begin{center}
\includegraphics[width=6.2cm]{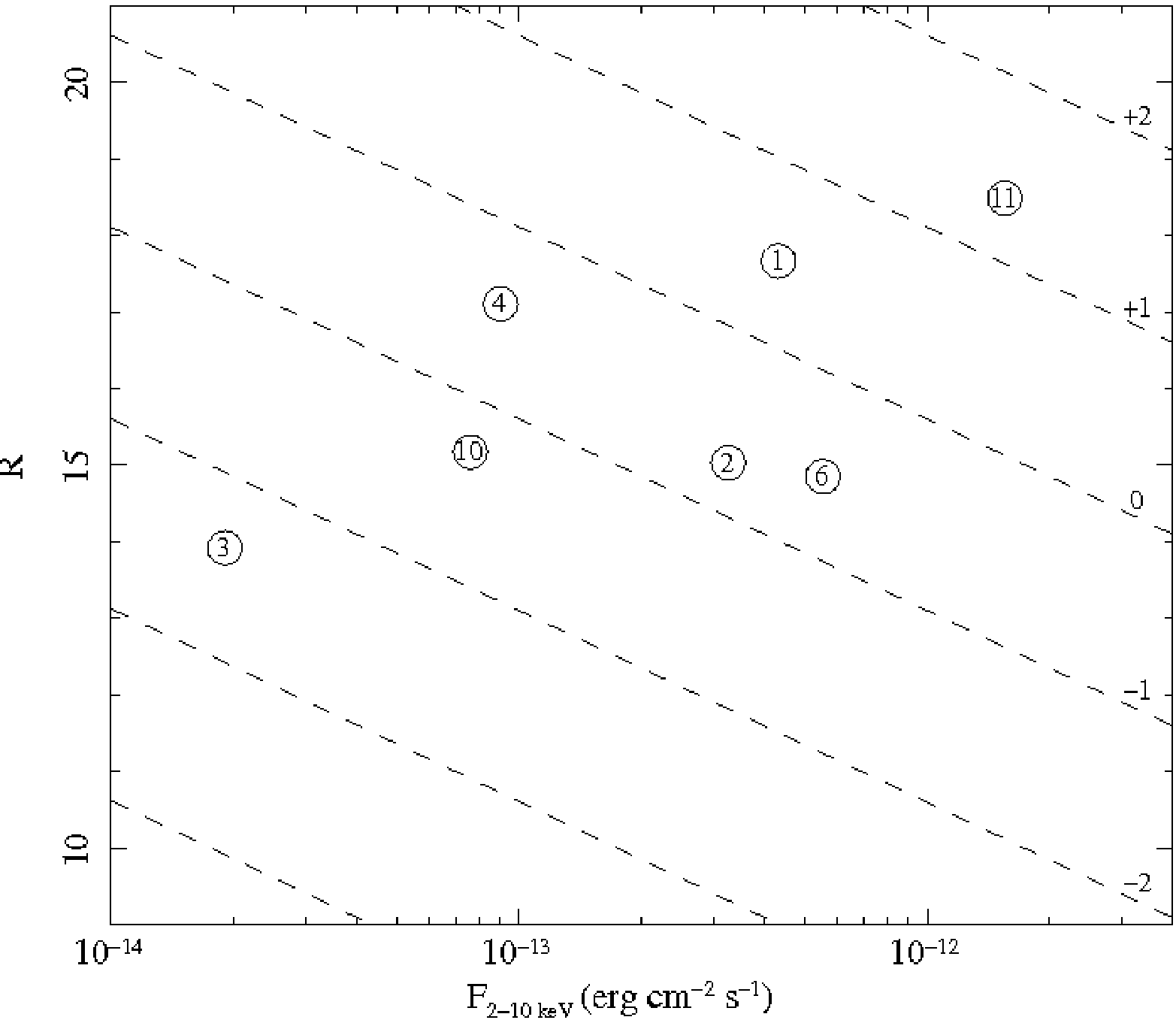}
\includegraphics[width=6.2cm]{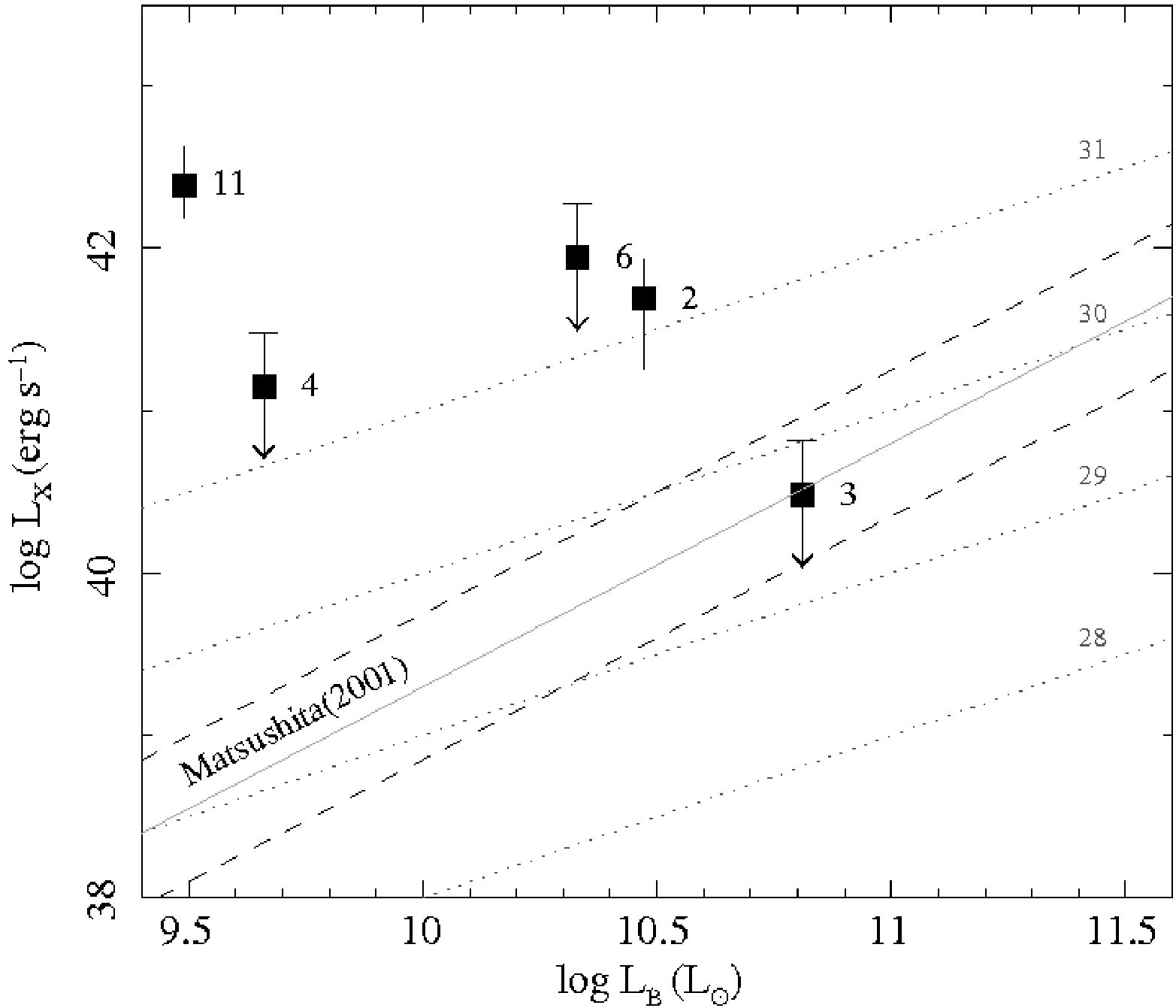}
   \end{center}
  \caption{Left panel: Optical $R$-band magnitude vs. $F_X$ for X-ray detected
    galaxies in our sample. Diagonal dotted lines represent lines of constant
    X-ray to optical flux ratio [i.e. log($f_X$/$f_R$) = +2, +1, 0, -1,
    -2]. Luminous AGNs generally have log($f_X$/$f_R$) $\ge$ -1.
    Right panel: L$_X$-L$_B$ relation from A2255 cluster field.
The gray dashed lines show the slopes for log (L$_X$/L$_B$) ratios of 28, 29, 30 and 31 for visual
aid. The solid line is the distribution function for X-ray compact galaxies
defined by Matsushita (2001) 
with 90$\%$ confidence limits (dashed lines).} 
\label{fig:R-Fx}
\end{figure}

In order to interpret the origin of this excess emission, we compared x-ray and optical properties. 
If the emission in optical wavelength is also elaveted with the same amount,
the x-ray excess would not be suprising. 
According to the standard picture of galaxy clusters, red, 
early-type galaxies with less star formation are longer
residents (Refs. \refcite{ell-04}; \refcite{kod-01}). 
Blue, later-type galaxies are more likely to locate at outer regions (e.g. Refs. \refcite{dre-80}, \refcite{whi-93}). 
In this study of A2255, the sources at larger radii (R $>$ 0.5 Mpc), the
populations from red cores to bluish outskirts were emphasized.
We calculated B and R magnitudes of the optical follow-up sources through our {\em  RTT-150} observations.
Figure \ref{fig:R-Fx} shows the comparisons of flux (lef panel) and luminosity (right panel) ratio values. 
The details are discussed in the following sections, \ref{fx/fr} and \ref{lx/lb}.

\subsection{\label{fx/fr}X-ray to optical fluxes - log ($f_X$/$f_R$) }
X-ray to optical flux ratio is a very helpful tool and provides important
information on the variety of X-ray sources (Ref. \refcite{mac-88}). 
The ratio is applied in many X-ray surveys for its simplicity and clarity 
(e.g. Refs. \refcite{kru-07}; \refcite{mai-02}). 
Figure \ref{fig:R-Fx} shows the $R$-band magnitudes (from {\em ANDOR}) versus hard
band flux for X-ray detected galaxies in our sample. 
We have illustrated several diagonal dashed lines to indicate constant X-ray to optical flux ratio 
values [log ($f_{X}/f_{R}$) = $+$2, $+$1, 0, $-$1, $-$2] in the figure
which are calculated by the Kron-Cousins $R$ filter transmission function as defined by Ref. \refcite{zom-90},
$\rm{log} ({\it f_X}/{\it f_R}) = \rm{log} ({\it f_X}) + 5.759 + {\it R}/{2.5}$.
Since the ratio is independent of distance, it is very useful to distinguish X-ray
source variation by impeding redshift uncertainties.
The normal galaxies and stars have log ($f_{X}/f_{R}$) $\le$ $-$2 and AGNs
show higher ratio values (Ref. \refcite{mai-02}).
Log ($f_{X}/f_{R}$) $\ge$ $-$1 is comprehended as clear indication of
unobscured AGN activity (e.g. Ref. \refcite{leh-07}).
A Large fraction of our sources spans in the ratio range of ($+$ 1 $\sim$ $-$ 2) which is typical for AGNs. 
Log ($f_{X}/f_{R}$) plot (Figure \ref{fig:R-Fx} left panel) demostrates that
majority of our samples are either a bright x-ray source or an AGN.
However, x-ray to optical flux ratio is insufficient for a definite explanation of source types.
We explore X-ray-to-optical luminosity ratios ($L_X$/$L_B$) in the next
section to explore possible AGN activity.

\subsection{\label{lx/lb}X-ray to optical luminosity relation ($L_X$/$L_B$)}
X-ray studies of local early-type galaxies have revelaed that the total X-ray
luminosity is associated with optical luminosity (e.g. Refs. \refcite{osu-01}; \refcite{elli-04}). 
Ref. \refcite{sar-06} reviews the nature of elliptical galaxies and explains how the interstellar gas is heated by high velocity
interactions of stars and kinetic energy produced by Type Ia supernovae. 
Consequently, the $L_X$/$L_B$ luminosity ratio demonstrates the stellar velocity dispersions of a definite galaxy ($L_X$) and the stellar mass
($L_B$) of that galaxy (Refs. \refcite{leh-07}; \refcite{mah-01}).
Ref. \refcite{can-87} defines X-ray to optical luminosity correlation as $L_X$ $\propto$ $L_B^{1.6-2.3}$ for optically bright sources. 
The correlation for optically faint galaxies are observed to be rather linear,
$L_X$ $\propto$ $L_B$ (e.g. Ref. \refcite{gil-04}).  

We plotted logarithm of the 2$-$10 keV luminosity
($L_X$) vs. the logarithm of the $B-$band luminosity ($L_B$), in Figure
\ref{fig:R-Fx} right panel. 
The dotted gray lines show log($L_X$/$L_B$) $=$ 28, 29, 30, 31 for visual aid. 
For comparison the average ratio distribution of early-type galaxies and 90\%
confidence limit reported by Ref. \refcite{mat-01} is represented in solid and dashed lines. 
A positive dispersion is evident in the plot. The majority of the sources have
log($L_X$/$L_B$) $>$ 31. One source (\#3) shows the properties of normal elliptical galaxy. 


The above mentioned $L_X$/$L_B$ plot (Figure \ref{fig:R-Fx} right panel) was
obtained only for the sources with optical counterparts. 
We also checked $L_X$/$L_B$ ratios for the sources with no optical follow-ups
(i.e. \#7, \#8 and \#9). 
Considering the limiting magnitute of $R$=18.48, optical luminosity values were assinged. 
Source \#7 is found to be in the range of normal ellipticals. 
And sources \#8 and \#9 have the ratio similar to that of either bright ellipticals or AGNs.    

\subsection{\label{notes} Notes on individual sources}

\textbf{XMMU J171359.7+640939} (\#1) and \textbf{XMMU J171139.1+641405} (\#10) are background objects. 
The X-ray spectra were best fitted by fairly large power-law indices ($\Gamma$ $>$ 2.2). 
Both sources are identified as QSOs, with $z$ = 1.363 (Ref. \refcite{sch-02}) and 
$z$ = 0.925 (Ref. \refcite{ric-04}), respectively. 
The EPIC-PN X-ray counts were insufficient for redshift determination.
And yet in Figure \ref{fig:R-Fx} both sources show strong deviation. 
In this study we verified that they both are QSOs and do not reside in A2255 gravitatinal potential. 

\textbf{XMMU J171342.3+640454} (\#2) is a soft X-ray source. 
About 60\% of the X-ray emission if detected below 2 keV.
The best fitted power-law model gives an index of $\Gamma$=1.29$^{+3.53}_{-1.05}$.
Chandra data was better fitted with 0.57 keV thermal emission of an elliptical galaxy (Ref. \refcite{dav-03}). 
However, in our survey the net source count was not of sufficient quality to rule out thermal emission.
The source is possibly interacting with the nearby ($\sim$ 20 kpc) spiral
galaxy (Fig. \ref{fig:optic}). 
The {\em TFOSC} spectrum shows a weak feature at 4034 $\rm \AA$ which is
assumed to be [O $_{\rm II}$] emission line (Fig. \ref{fig:optic}). 
The determined redshift for this source is $z$ = 0.0824. 
It is comparable to the value of Ref. \refcite{zwi-71} ($z$ = 0.0847), formerly known as ZwCl 1710.4+6401 40.
The X-ray and optical properties are consistent with that of Chandra (Ref. \refcite{dav-03}).
The source has log ($f_{X}/f_{R}$) $>$ $-$1 and $L_X$/$L_B$ = 31.22 (see Fig. \ref{fig:R-Fx}).
These values are higher than those of nornal elliptical or starburst galaxies.
Based on the above results, the source is very bright galaxy, if not classified as an AGN. 

A new redshift value is determined for \textbf{XMMU J171325.4+641000} (\#3).
{\em TFOSC} spectrum has an [O $_{\rm II}$] emission line at 4029.3 {\rm $\AA$}. 
The redshift we measure is $z$ = 0.0811 which is the first distance calculation for the source.
The morphology shows signs of elongation in sout-east direction.
The X-ray emission is dominated in soft energies and L$_X$ $=$ 3.0 $\times$ 10$^{40}$ ergs s$^{-1}$.
The object negatively deviates in the flux and luminosity ratios (see Fig. \ref{fig:R-Fx}). 
It has log ($f_{X}/f_{R}$) = -2.4 and $L_X$/$L_B$ = 29.67 which is expected from a normal galaxy.
Being the faintest X-ray source in our sample, fairly bright in $B-$band 
and lacking of strong AGN features, we identify this object as a normal galaxy. 
The source is not buried into the extend ICM X-ray emission as has been shown
Figure \ref{fig:A2255_with_sources}. 
It is probably approaching to the cluster potential in the FOV direction. 

Sources \textbf{XMMU J171236.2+640035} (\#4) is fairly bright.
{\em TFOCS} spectrum showed [O $_{\rm II}$] line at $\sim$ 4056 $\rm \AA$
(Fig. \ref{fig:optic}). The obtained redshift is $z$ = 0.0883.
The x-ray and optical peaks do not coincide. The shift is about 4 arcsec
($\sim$ 6.5 kpc) (See Fig. \ref{fig:optic}). 
Considering x-ray and optical positional errors, this distance can be presumed in the confidence range.
The calculated redshift value suggests that the source is merging to cluster from behind in the south-west direction.

{\em TFOSC} spectrum of source \#5 (\textbf{XMMU J171220.9+641007}) shows M star-like properties. 
Strong TiO absorbtion is evident in the optical spectrum of Figure \ref{fig:star}, which is strong in M-stars. 
The absorption band appears near 7590 $\rm \AA$. 
The X-ray spectrum of the source is fairly hard ($\Gamma$ $\sim$ 1.3) which
is probably caused by magnetic activity (Ref. \refcite{bra-05}). 
X-ray flux is obtained to be 5.32 $\times$ 10$^{-13}$ ergs cm$^{-2}$ s$^{-1}$. 

\textbf{XMMU J171216.2+640210} (\#6) is the second brightest X-ray source in our survey. 
X-ray properties is also studied by Ref. \refcite{dav-03} with {\em Chandra}. 
Our best fitted spectra result and calculated luminosity with {\em XMM} is
consistent with that of {\em Chandra} (Ref. \refcite{dav-03}). 
The galaxy is also a powerful head-tail radio source (Ref. \refcite{fer-97}).
The emission line at 4373.6 $\rm \AA$ in {\em TFOSC} spectrum is evident. 
If this line is associated to S $_{\rm II}$ (rest frame $\lambda$4072), 
our calculated redshift value of $z$ = 0.0741 is comparable to the $z$ = 0.0714 value of Ref. \refcite{owe-95}. 
The optical spectrum shows emission features of [O $_{\rm II}$] (rest frame
$\lambda$3727) and [H$\alpha$] at 7121.5 $\rm \AA$ (see Figure \ref{fig:optic}). 
[H$\alpha$] line is seen from galaxies classified as star-forming.
The Ca $_{\rm II}$ H and K and CH around 4665 $\rm \AA$ (rest frame 4280-4300
$\rm \AA$) absorption bands are also weakly present. 
The weak absorption features at 7053.4 $\rm \AA$ is assumed to be Ca $_{\rm II}$ + Ba $_{\rm II}$ blend.
These match the redshift to be $z$ = 0.0851 and place the source closer to the
cluster potential well, which is very tempting.
A consistent redshift value ($z$ = 0.0714) is observed, but $z$ = 0.0851 as indicated by weaker lines is not totally ruled out. 
The $f_{X}/f_{R}$ and the $L_X$/$L_B$ values are consistent with that of AGNs. 
Ref. \refcite{col-02} studied luminous X-ray source that are not AGNs at $L_X$ $\sim$ 10$^{41}$ ergs s$^{-1}$. 
Therefore, source \#6 appears to be a very luminous galaxy, but may also be an AGN as noted by
Ref. \refcite{dav-03} as well.

The sources with no optical follow-ups (\#7, \#8 and \#9) were bright in X-rays (see Table \ref{table3}). 
Considering the confidence range of the X-ray luminosities, \textbf{XMMU J171159.6+635938 (\#7)} can be explained by normal ellipticals. 
The X-ray emission from \textbf{XMMU J171148.1+635601(\#8)} and \textbf{XMMU J171148.7+640534 (\#9)}
is contaminated by unresolved discrete sources in these galaxies.
X-ray spectrum of \textbf{XMMU J171123.7+641657 (\#11)} is studied in detail for the very first time in this study.
The $L_X$/$L_B$ study (see Fig. \ref{fig:optic}) indicates that source \#11 may be an AGN. 
It is the brightest X-ray source of our sample, but the faintest in optical band (see. Fig \ref{source_spectrum}).
if sources \#8, \#9 and  \#11 really possess active nuclei, the possible mechanism that
would obscure the optical emission from sources is very intriguing.
One of the possible explanations would be X-ray bright optically normal galaxies (XBONGs).
XBONGs are thought to be luminous AGNs (Refs. \refcite{com-02}; \refcite{yua-04}). 
This is also an ideal solution for our result, especially regarding our
luminosity value of $L_X$ = 2.5 $\times$ 10$^{42}$ erg s$^{-1}$ for source \#11.

The recent studies show that the X-ray sources are not randomly distributed
within clusters (Refs. \refcite{mar-02}, \refcite{joh-03}). 
The brighter sources are likely to reside at the outskirts, while faint ones
are closer to the core (R $<$ 1Mpc). 
Activating or deactivating x-ray sources are discussed with possible AGN
fueling and quenching mechanisms (Refs. \refcite{mar-06}, \refcite{hud-06}).
Considering the luminosity range of our sample (10$^{40.8}$ $<$ $L_X$ $<$
10$^{42.39}$ erg s$^{-1}$), the sources from A2255 outskirts are likely to be
undergoing a similar physical processes. The cluster environments are the
places with high probability of collisions and close passages. 
Particularly, the outskirts are dynamically complex due to new arrivals.
Normally galaxies are inbalance within itself, particularly in terms of accreeting matter.
When a galaxy begins to experience cluster gravitational potential and
encounters high density ICM, the balances are then disturbed. 
The stability between accreeting system is rearranged. 
Depending the mass of the compact object, new LMXBs and AGNs might be formed
or they get brighter as more matter accretes. 
It is known that, if not all, galaxies host a black hole (BH) at its center. 
Most of them are not energetic, waiting for fuel to awake. 
The dynamical high density environment of cluster outskirts may trigger these kinds of
starved BHs, which increases the X-ray source density in cluster outskirts.
This explains excess of X-ray emission and crowd of x-ray sources in A2255 outskirts.
The infall induced nuclear activities is the suggested solution for our results. 

\section{Conclusions}
With {\em XMM} archival data we detected 11 point sources at the outskirts of A2255.
Optical follow-up sources are observed by {\em RTT-150} optical telescope.
X-ray spectra for several sources (\#7, \#8, \#9 and \#11) were studied for the very first time.
The results for source \#2, \#4 and \#6 are found to be in reasonably good
aggreement with their earlier studies (Refs. \refcite{dav-03}, \refcite{yua-04}).
Source \#5 identified to be an M-type star with strong TiO absorption.
The redshift value of $z$ = 0.0811 found for source \#3 through our 
{\em TFOSC} optical observations. 
The sources \#1 and \#10 verified to be possible QSOs. 
Based on log($N$)-log($S$) study, an elavated X-ray emission and/or higher
source population is evident from the cluster outskirt. 
The source population is calculated to be 4 times higher than the field at the
flux value of $F_X$ $\sim$ $10^{-13}$ erg cm$^{-2}$ s$^{-1}$.
We suggest that X-ray emission is triggered by either increased accretion
onto LMXBs, fueling of AGNs and awakening of BHs in the cluster outskirts.

\section*{Acknowledgement}
This work is financially supported by a Grant-in-Aid by Academic R\&D Funding System
contract No. 106T310, through The Scientific \& Technological Research Council of Turkey (TUBITAK). 
M.H. is supported by TUBITAK Post-Doctoral Fellowship.
The authors also acknowledge the partial support by
Bo\u{g}azi\c{c}i University Research Foundation via code number 06HB301.
We would like thank Dr. Graziella Branduardi-Raymont for her useful suggestions and comments.
This research made use of data obtained through the XMM-Newton,
an ESA Science Mission with instruments and contributions directly funded by
ESA member states and the US (NASA).

\end{document}